\documentclass[aps,pra,amsmath,amssymbol,stfloats,twocolumn]{revtex4-1}
\usepackage{amsfonts, bbm}
\usepackage{amsbsy,latexsym,amsthm,mathrsfs,bbold}

\usepackage{amssymb}
\usepackage{hyperref}
\usepackage{epsfig,graphics,graphicx}
\usepackage{color}
\usepackage{enumerate}

\newcommand{\ket}[1]{\vert#1\rangle}
\newcommand{\bra}[1]{\langle#1\vert}
\newcommand{\mean}[1]{\langle#1\rangle}
\newcommand{\braket}[2]{\langle#1\vert#2\rangle}
\newcommand{\ketbra}[2]{\vert#1\rangle\langle#2\vert}
\newcommand{\ketbrad}[1]{\vert#1\rangle\langle#1\vert}
\newcommand{\id}{\openone}
\newcommand{\mse}{\sigma^{2}}
\newcommand{\msej}{\sigma^{2}_{j}}

\newcommand{\tr}[1]{\mathrm{tr}(#1)}

\newcommand{\expo}[1]{\mathrm{e}^{#1}}

\newcommand{\ml}{\ell}

\begin{document}
\title{Probabilistic metrology or how some measurement outcomes render ultra-precise estimates}
%\title{Ultimate precision bounds from probabilistic quantum metrology}
\author{J. Calsamiglia, B. Gendra, R. Mu\~{n}oz-Tapia, and E. Bagan}
\address{F\'{i}sica Te\`{o}rica: Informaci\'{o} i Fen\`{o}mens Qu\`antics, Departament de F\'{\i}sica, Universitat Aut\`{o}noma de Barcelona, 08193 Bellaterra (Barcelona), Spain}

\begin{abstract}
%{\color{blue}Quantum-enhanced measurements exploit quantum mechanical effects to provide precise estimates of physical variables for use in advanced technologies, such as frequency calibration of atomic clocks, gravitational waves detection, and biosensing. Quantum metrology studies the fundamental limits in the estimation precision given a certain amount of resources (e.g. the number of probe systems) and restrictions (e.g. limited interaction time, or coping with unavoidable presence of noise). }

We show on theoretical grounds that, even in the presence of noise, probabilistic measurement strategies (which have a certain probability of failure or abstention) can provide, upon a heralded successful outcome, estimates with a precision that exceeds the deterministic bounds for the average precision. This  establishes a new ultimate bound on the phase estimation precision of particular measurement outcomes (or sequence of outcomes). For probe systems subject to local dephasing, we quantify such precision limit as a function of the probability of failure that can be tolerated. 
Our results show that the possibility of abstaining can set back the detrimental effects of noise.
\end{abstract}

\maketitle

\section{Introduction}
Quantum-enhanced precision measurements and sensors are some of the most disruptive quantum technologies \cite{dowling_quantum_2003}, with applications  across various disciplines, e.g.,  optical communications \cite{slavik_all-optical_2010,chen_optical_2012},  cryptography \cite{inoue_differential_2002},  brain and heart  medical diagnosis  via atomic magnetometry~\cite{budker_optical_2007,napolitano_interaction-based_2011}, biological measurements \cite{taylor_biological_2013,crespi_measuring_2012}, and are critical in gravitational-wave detectors \cite{caves_quantum-mechanical_1981,collaboration_gravitational_2011} and GPS and other current technologies that rely on atomic clocks   \cite{bollinger_optimal_1996,huelga_improvement_1997,wineland_spin_1992,komar_quantum_2014}.

The above are examples of the so-called quantum metrology problems. In broad terms a metrology problem  can be cast as a four step process: the preparation of a probe, its controlled evolution that imprints in the probe the (continuous) unknown parameter to be estimated,  a measurement on the modified probe, and a final data-processing to obtain the value of the unknown parameter.  The accuracy of the estimation is limited by the experimental imperfections and, ultimately, by the noise  inherent in any quantum measurement. Classically, one can reduce the effects of noise in a given setup by repeating the experiment  on $n$ independent preparations of the probe~\cite{giovannetti_quantum_2006}. The uncertainty of the estimation is thereby reduced by a factor~$n^{{-1/2}}$ (the so-called standard quantum limit, SQL). However, in a fully quantum mechanical setting, the possibility of  using  entangling operations in the preparation and in the measurement steps gives rise to an uncertainty that scales as~$n^{{-1}}$ (the so-called Heisenberg limit or scaling).

Recent experimental advances that enable an unprecedented control of diverse optical and condensed matter  systems at a quantum level make quantum metrology an extremely timely field of research \cite{giovannetti_advances_2011}. In the last years, the agenda of quantum-enhanced metrology has been put under scrutiny by a number of results~\mbox{\cite{demkowicz-dobrzanski_elusive_2012,escher_general_2011,knysh_scaling_2011,aspachs_phase_2009,huelga_improvement_1997,Jarzyna_2015}} that show that under  quite generic (local, uncorrelated and markovian) experimental noise the  quantum enhancement  amounts  to a constant factor rather than a quadratic improvement.  The field has revamped in search for alternative schemes that push forward the limits and circumvent or diminish the detrimental effect of noise. This has entailed the study of particular systems with non-trivial noise-models~\cite{chaves_noisy_2013,chin_quantum_2012,jeske_quantum_2013,ostermann_protected_2013,szankowski_parameter_2012}, and  non-linear interactions \cite{boixo_generalized_2007,napolitano_interaction-based_2011} that enable quantum error-correction codes~\cite{macchiavello_2000,dur_improved_2014,kessler_quantum_2014,arrad_increasing_2014}.
 
Most quantum metrology schemes found in the literature  and their corresponding bounds are deterministic. That is, these schemes are optimized in order to provide a valid estimate for each possible measurement outcome, in such a way that the average precision is maximized.  
Recently it has been shown that for a fixed probe state, and in the absence of noise, the precision of some particular (favorable) outcomes can be greatly enhanced, well beyond the limits set for deterministic strategies~\mbox{\cite{fiurasek_optimal_2006,gendra_quantum_2013,gendra_optimal_2013,marek_optimal_2013}}. The possibility to post-select, i.e., to abstain from providing an estimate some times, can even change the uncertainty from SQL to Heisenberg scaling.  It has also been shown that the limit on the precision of these probabilistic metrology strategies agrees with that found for deterministic strategies when the optimization over probe states is performed. So,  for pure states, probabilistic metrology can compensate a bad choice of probe state, or in other words, it can attain the optimal precision bounds in situations where the probe state is a given. 

Here we study the performance of probabilistic metrology in the presence of noise. We will show that probabilistic metrology can substantially lessen the effects of local dephasing noise, although not enough to overcome the infamous loss of asymptotic Heisenberg scaling~\cite{escher_general_2011,demkowicz-dobrzanski_elusive_2012}. In addition, and in contrast to the noiseless ideal case, the ultimate precision bounds for probabilistic metrology will be shown to exceed those attained by deterministic strategies optimized over probe states.

To put these results into context, we recall that in most quantum metrology schemes the probe is a composite made up of a large number of elementary quantum systems [1-35]. We then envisage the following situation:
%To put our results into context and before stating our framework in next section, let us invoke the following scenario: 
An ensemble of fifty-thousand  two-level atoms has been prepared in a known probe state $\rho$ and is awaiting for  a signal coming from a supernova, such as a gravitational wave or some byproduct of a gamma burst. The experiment is designed in a way that the signal {will leave} an imprint on the state of the atoms, $\rho\to\rho_{\theta}$,  that  will depend on the value of some relevant physical parameter $\theta$ of the signal. The experimenter will perform a measurement on the atoms and will try to infer from the outcome the unknown value of~$\theta$. The experimental set-up is perfectly calibrated and characterized.  At a certain time the long-awaited event occurs. Conditional on the obtained measurement outcome, the experimentalist reports a value of  $\theta=2.23$ with a  mean square error of $\sigma^2=10^{-7}$.  How should the community react if  the reported error is smaller than the (deterministic) bounds found in current literature,   e.g. ,~$\mean{\sigma^2}<1/n=2 \cdot10^{-4}$?

The direct answer is that the community should celebrate the result without reservation. The error obtained in a single outcome can be smaller than the corresponding limits found in the literature, which are based on bounds on the average error over all outcomes.  It is no surprise that the precision depends on what particular observation one happens to obtain: some observations are better, more  informative, than others. The apparent {contradiction} disappears when one realizes that ultra-precise outcomes can only occur if {infra-precise~outcomes also} exist  (so as to respect the deterministic bounds).

Here {we} show that by a suitable choice of measurement it is possible to obtain \emph{ultra-precise}  outcomes, whose precision is {still} limited, but goes beyond the (average) bounds established for deterministic quantum metrology protocols. We also show that such ultra-precise outcomes can only occur with a small probability. 

In a scenario where each outcome can have a different precision, the criterion for optimality is by no means unique.  In a classical setting there is no compromising choice to be made. One can find the optimal estimator and precision for each measurement outcome independently. However, in the quantum case one needs to fix the POVM through some criteria. The ultimate quantum limit is obtained by choosing a POVM that produces an outcome with the highest possible precision;  deterministic protocols are optimized in order to produce the highest possible precision on average (over all possible outcomes). The protocols that we study here under the name of probabilistic quantum metrology interpolate between these two  cases, by optimizing the precision of successful outcomes with the constrain that they occur with some prescribed probability.

Understanding the power of probabilistic operations in  general quantum tasks is a highly non-trivial and relevant problem. %in quantum information sciences. 
Probabilistic operations introduce through normalization a very particular non-linearity that is in stark contrast with the linearity of quantum deterministic operations.  Many no-go theorems stem from linearity %of quantum mechanics, 
and probabilistic operations might revoke them, turning the once-thought impossible into possible. Notable examples include unambiguous state discrimination, whereby non-orthogonal states can be distinguished with no error~\mbox{\cite{Ivanovic}} (see \cite{fiurasek_optimal_2003,bagan_optimal_2012} for generalizations); probabilistic cloning \cite{cloning}
%
%A deterministic protocol that minimizes the average probability of error, produces two outcomes (one for every state). Each outcome has a given probability to give the incorrect answer, and at first sight it seems impossible to reduce  it beyond its optimal value. However, if a third outcome (abstention) is included, it is possible reduce the probability of an erroneous identification down to zero \cite{}, what is known as unambiguos state discrimination.} 
%Similarly, the limitations imposed by the no-go theorem of realizing Bell measurements by linear-optical elements \cite{lutkenhaus_bell_1999} was overcome by 
the KLM scheme~\cite{knill_scheme_2001}, whereby Bell measurements can be realized by linear optical elements~\cite{lutkenhaus_bell_1999};
%In quantum computing, most quantum algorithms require a certain (bounded) error probability. Exact algorithms that  provide a quantum speed-up and output the correct answer with certainty  are hard to come by \cite{ambainis_exact_2014,montanaro_exact_2013}.  ; 
also,  related to the current work we find  probabilistic amplification \cite{ferreyrol_implementation_2010, xiang_entanglement-enhanced_2011,zavattaa._high-fidelity_2011,kocsis_heralded_2013,chiribella_optimal_2013,pandey_2013} and
weak-value amplification~\mbox{\cite{dressel_colloquium:_2014,hosten_observation_2008,brunner_measuring_2010,zilberberg_charge_2011,jordan_technical_2014}}.

Although {the probability of attaining the ultimate bounds} is often small, at a fundamental level it is important to distinguish between ultimate versus {\em de facto} quantum limits. No matter how unlikely an event is, once it occurs it is a certainty; and certainties cannot violate ultimate bounds \cite{calsamiglia_quantum_2014}.
%: faster than light signaling is impossible, winning the lottery (having bought the ticket) is not likely, but perfectly plausible. In the same way tunneling is highly improbable, but it can lead to spectacular or catastrophic consequences. 
This fundamental distinction has also motivated the definition of a complexity class in quantum computing \cite{aaronson_quantum_2005}. All in all, post-selection can be considered a resource per se in quantum information tasks, and this paper is devoted to the study of its power for metrological tasks in realistic noisy scenarios.

Our work becomes particularly relevant in applications  where:
a) {\em There are high demands on precision}.  We are already at a stage {where} quantum metrology is required to push the limits of precision. Hence it might well be that {a} known optimal deterministic protocol, {e.g.,} the phase covariant measurement, fails to provide the precision required for a specific task, {whereas a probabilistic scheme does not}. There are tasks for which having an estimate below a certain precision is {\em at least} as bad as having no estimate at all. For instance, when locating a tumor in radiation therapy, or some deeply buried magnetized material for its extraction, missing the true position of the target by more than certain threshold value can have disastrous consequences. b) {\em Resources are fixed}. {As in the first example above, it might be impossible to repeat an experiment} (for a given instance of the unknown parameter). 

In real life applications, we care about the precision {attainable} in absolute terms. We wish to add consistent error bars to our estimates,  not just demonstrate a particular scaling of the uncertainty as the number of resources increases. Our results hold both for finite and asymptotically large number of resources. Fixing the number of resources is not only necessary to state optimality in a {clear-cut way. In} many situations the limitations on the {available resources} are  patent. It might be a given, {as in the measurement of the magnetization of a particular magnet; or in the measurement of a parameter that changes rapidly with time (a requirement in some feedback schemes)}. Last but not least, there are situations in which the experiment is impossible to reproduce because the event under study is uncontrollable, {e.g.,} a supernova or the arrival of a gravitational wave. {It is precisely in observational astronomy where} the (classical) bayesian approach to statistical inference, on which our analysis relies, is widely used in current studies (see for instance \cite{benitez_2000}). 

%From a technical point of view, we introduce a novel technique to derive the optimal probabilistic measurement and its precision that exploits the formal equivalence with the problem of computing the ground state and energy of a particle in a one-dimensional potential box, with some boundary conditions that depend on the strength of the noise and on the initial probe state.

 This paper is organized as follows. Section II describes the theoretical framework of this work. We state the general working assumptions and discuss the statistical approach(es) used throughout the paper. Section III contains the core findings of our work. To ease the presentation, we have divided it into different subsections that contain different results.  We give general expressions for the uncertainty and probability of success for covariant measurements, closed expressions for asymptotically large number of resources as well as the optimal probe states and the ultimate precision bounds. We also address the 
 issue of the information left in the system after a discarded event.
%Our results are given in Section III, which is divided in different subsections: In A, the general expressions for the precision and the probability of success for covariant measurements are given. They are further simplified in B for permutation-invariant probe states.
%Subsection C exploits a formal analogy with the problem of a particle in a 1-D box to compute the precision for an asymptotically large number of resources.
%Closed asymptotic expressions for multi-copy probe states are given in D. They are compared  with the exact results for finite number of resources in E. We obtain the optimal probe states that reach the best precision for a given success probability, and thereby and establish the ultimate quantum precision bound in F. Finally, Subsection G  shows that, although the probabilistic scheme prioritizes the precision of the favorable events, how some information on the unknown parameter can be in principle be recovered from discarded events. 
In the last section we state our main conclusions and discuss possible implementations of our scheme. More technical results are presented in the appendices. In particular, \ref{subs:notation} introduces specific notation that is used in the derivation of some of these results.

\section{Framework \label{sec:framework}}

In this section we introduce our framework in detail. We  set out our physical assumptions and goals, and discuss why, in view of these assumptions, the Bayesian formulation suits our purpose better, while also allowing for a straightforward extension to the probabilistic case. 
%
%Quantum metrology %, as any statistical inference problem, can be approached from a frequentist or Bayesian perspective. Though 
%has traditionally been approached from a frequentist viewpoint, however, it does not apply to our framework, as will be apparent from our discussion. 
% the Bayesian formulation has also been used and is gaining pace in the quantum context, where the choice is no longer a matter of taste and convenience as will be argued. 
Alternatively, the minimax approach, or worst case scenario, is also well  suited.  For the problem at hand, the later is shown to be equivalent to the Bayesian formulation. The relationship with the frequentist approach is discussed at the end of the section.

%Quantum metrology has been traditionally approached from a frequentist perspective. This is in contrast with other problems in quantum state estimation, where both, frequentist and Bayesian points of view are used (roughly) on an equal footing,  or quantum state discrimination, where~the Bayesian viewpoint is predominant. In most cases, the choice is  largely a matter of convenience or taste. Recently, there has been some controversy about this issue in the literature. We, thus, feel compelled to introduce our framework in detail. In this section, we state our physical assumptions, which in turn explain why we stick to the Bayesian viewpoint. Alternatively, the minimax approach, or worst case scenario, is also well fitted to our framework. For the problem at hand, the later is shown to be equivalent to the Bayesian approach.

% Metrology, and in particular quantum metrology, can be approached from different angles strongly influenced by the particular practical application or fundamental question that is under study.
%Let us introduce the scenario where our results are framed by spelling out our general working hypothesis:

%com la transició al cas probabilistic a causat certa controversio ..let us remind the reader whta is the bayesian approach about..

Our framework consists of the following:

\begin{enumerate}[a)]

\item {\em A model}. We assume there exists a model that formalizes the ``state of the world" and the measurement.  In our case the former is given by the quantum state~$\rho_{\theta} %=U_\theta \rho\, U_{\theta}^{\dagger}
$ (see Sec.~\ref{sec:results}) of a physical system of interest, where~$\theta$ is a real parameter. It can, of course, take into account noise sources and other %appropriately 
experimental imperfections. %The  state of the world, parametrized by $\theta$, represents the physical system of interest. In our case it is given by a quantum state~$\rho_{\theta}=U_\theta \rho U_{\theta}^{\dagger}$ (see Section \ref{sec:results}). 
The measurement outcomes and the state of the world %, labeled by~$\theta$, 
are related through a measurement model (Born's rule in our case) that gives the probability distribution of the outcomes for a given $\theta$.
The true value of~$\theta$ is assumed to be unknown, while the rest of the model parameters are known with high precision. Our goal is to infer the value of~$\theta$ from the observed measurement outcome. 

%\item {\em A model}. We assume there exists a model (determined by the laws of quantum mechanics) that formalizes the ``state of the world" and the measurement.  In our case and it can of course include noise sources and other appropriately characterized  experimental imperfections. The  state of the world, parametrized by $\theta$, represents the physical system of interest. In our case it is given by a quantum state $\rho_{\theta}=U_\theta \rho U_{\theta}^{\dagger}$ (see Section \ref{sec:results}). The observations, or measurement outcomes, are related to the state of the world $\rho_{\theta}$ through a measurement model (Born's rule) that relates the probability distribution of the outcomes to $\theta$.
%The true value of the parameter  $\theta$ is unknown and it is the aim of the scheme to infere it from the measurement outcome. 

\item {\em A fixed amount of resources}. We view the size of the probe state $\rho_\theta$ as the amount of resources of the problem. More precisely, we quantify the amount of resources by the number $n$ of qubits that the estate $\rho_\theta$ describes. Accordingly, if an experiment consists of repeating a measurement on independent copies of a system of~$N$ qubits a number $\nu$ of times, the total number of resources used is $n=\nu N$.

\item {\em An optimization over measurements}. %We want to establish ultimate bounds on the precision that can be reached with given resources. In a quantum setting, the relation between the state of the world $\rho_{\theta}$ and the outcome probabilities, crucially depends on the choice of measurement.   The optimization has to be done over 
To establish ultimate bounds on the precision that can be achieved with given resources, we optimize over all measurements, 
the most general being a {\em single collective} measurement on the whole available resources, i.e., on~$\rho_\theta$. 
Hence, in our framework the inference protocols are single-shot. Namely, in each instance of the problem the state of the world is labeled by a particular (unknown) value of~$\theta$, and the measurement returns a single outcome $\chi$ out of the various possible outcomes of the measurement. Based on~$\chi$, an estimate~$\theta_{\chi}$ of the true value of~$\theta$ is produced.  

\end{enumerate}

Note that this characterization is fully general and includes strategies where each qubit or subsystem is measured individually, in which case every collective outcome is labelled by a sequence of individual measurement output labels. %Note also that without loss of generality we can label the measurement outcomes  by the value $\hat{\theta}$ of the corresponding estimate (see \ref{sec:results}).
%Without loss of generality, we can label the measurement outcomes by the value $\hat{\theta}$ of the corresponding estimate (see Sec.~\ref{sec:results}).

\begin{enumerate}[d)]

%\item {\em An optimization}. We want to establish ultimate bounds on the precision that can be reached with given resources. In a quantum setting, the relation between the state of the world $\rho_{\theta}$ and the outcome probabilities, crucially depends on the choice of measurement.   The optimization has to be done over the most general measurement, i.e. a single collective measurement acting on the resource state. Hence, in our framework the protocol is single-shot: in each instance of the problem the system of the world is in a particular (unknown) value $\theta$, and the measurement returns an outcome $\chi$---a single outcome out of many possible outcomes--- based on which an estimate $\theta_{\chi}$ of the true parameter $\theta$ can be provided.  Note that this characterization includes all possible measurement strategies, including those where each subsystem is measured individually. In that case every collective outcome will be labelled by a sequence of local measurement results. Note also that without loss of generality we can label the measurement outcomes  by the value $\hat{\theta}$ of the corresponding estimate (see \ref{sec:results}).

\item {\em A report of precision}.  %Any quantitative measure of the precision, i.e. the performance of a quantum metrological scheme, has to accommodate the above points. 
After a successful completion, the scheme should return an estimate $\hat{\theta}$ of the true parameter $\theta$, together with suitable error bars.  Error bars are essential in any scientific or technological discipline, as they quantify the confidence one should place in conclusions drawn from existing data. Such an assesment of the precision should be quantified bearing in mind that the whole set-up is single-shot, i.e., the experiment will not be necessarily repeated, and that the true value of the parameter is unknown. Hence, the precision assessment can only be based on the measurement outcome and on the precise knowledge of the model in~a).

%\item {\em A report of precision}.  Any quantitative measure of the precision, i.e. the performance of a quantum metrological scheme, has to accommodate the above points. After a successful realization, the scheme should return an estimate $\hat{\theta}$ of the true parameter $\theta$, together with an assessment of the precision associated to that result, an error bar so to speak. The precision should inform the experimenter about the accuracy of the estimate, bearing in mind that the whole set-up is single-shot, i.e. the experiment will not be necessarily repeated, and that the true value of the parameter is actually unknown to the experimenter. That is, the precision assessment can only based on the measurement outcome and on the precise knowledge of the model ---point a) above. 
%

 \end{enumerate}
 
%To have a concrete example in mind let us consider  the following hypothetical metrology experiment: An ensemble of thousands of atoms has been prepared in a known probe state $\rho$ awaiting for the detection of  a signal coming from a supernova such as a gravitational wave or some byproduct of a gamma burst (?). The experiment is designed in a way that the signal leaves an imprint on the state of the atoms $\rho_{\theta}$  that depends on some relevant physical property $\theta$ of the signal. The experimenter  performs a measurement on the atoms and tries to infer the value for unknown parameter $\theta$. Suitable error bars are also to be provided to know what confidence should be placed in conclusions drawn from the data that actually exists. 

 To carry out c) and d) we need to introduce a so-called cost or loss function~$\ml(\theta,\hat\theta)$ that quantifies how well our estimate $\hat\theta$ agrees with the true value of $\theta$. There are a priory infinitely many such functions, but two common choices in metrology are the quadratic loss function \mbox{$\ml_{\rm q}(\theta,\hat\theta)=(\theta-\hat\theta)^{2}$}, and 
$\ml_{\rm p}(\theta,\hat\theta)=4\sin^2[(\theta-\hat\theta)/2]$ if $\theta$ belongs to a periodic domain. Note that they are equivalent to leading order in $\hat\theta-\theta$, when the estimate approaches the true value, \mbox{$\hat\theta\approx \theta$}.

The Bayesian formulation offers a very natural and rigorous way to assign a quantitative precision measure to a particular outcome $\hat{\theta}$. In this formulation the unknown parameter $\theta$ is treated as a random variable and it is assigned a (prior) probability $\pi(\theta)$. This probability reflects the knowledge we have on the state of the world prior to~the measurement. After performing the measurement, the observed outcome and our knowledge of the model is used to update the prior $\pi(\theta)$ to the posterior probability distribution $p(\theta | \hat{\theta})$. Using Bayes' rule, we can write $p(\theta | \hat{\theta})=p(\hat\theta |\theta)\pi(\theta)/p(\hat{\theta})$, with $p(\hat{\theta})= \int\! {\rm d} \theta  \,{p(\hat\theta |\theta)\pi(\theta)}$.  Then the uncertainty of an outcome $\hat\theta$ is defined as
\begin{equation}
\label{Loss-theta} 
L_{\hat{\theta}}   =  \int\! {\rm d}\theta \,    p(\theta | \hat{\theta})\,  \ml(\theta,\hat\theta) .
\end{equation} 
 Thus, $L_{\hat{\theta}}$ quantifies how the unknown value $\theta$ is scattered around its estimate $\hat\theta$, in the light of the information gathered by the measurement.

In general, the various outcomes of a given measurement have different precision.
%A given measurement can produce different outcomes, each of them with a possibly different precision $L_{\chi}$. 
Hence, to quantify the overall performance of a metrology scheme by a single figure of merit we take the average uncertainty over all possible measurement outcomes,
\begin{equation}
\label{Loss-expected} 
L =\int\! {\rm d} \hat\theta\, p(\hat\theta) \,L_{\hat\theta}=\int\! {\rm d} \hat\theta \int\! {\rm d} \theta\, p(\theta ,\hat \theta)  \,\ml(\theta,\hat\theta),
\end{equation}
%
%\begin{eqnarray}
%L &=&\sum_{\chi}p(\chi) L_{\chi}= \sum_{\chi} \int\! {\rm d}\theta  p(\chi | \theta)  \pi(\theta)   f(\theta,\theta_{\chi})=\nonumber\\
%&=&\sum_{\chi} \int\! {\rm d} \theta p(\chi , \theta)  f(\theta,\theta_{\chi}).
%\end{eqnarray}
%{\color{red}where $p(\theta,\hat\theta)$ is the joint probability of the unknown parameter taking the value $\theta$ and the measurement outcome being $\hat\theta$.}
 where %$p(\theta,\hat\theta)$ is 
the joint probability is $p(\theta,\hat\theta)=p(\hat\theta|\theta)\,\pi(\theta)$.
Eq.~(\ref{Loss-expected}) is the expected loss $L$,  given by the average over all~possible values of the unknown parameter $\theta$ and~over~all measurement outcomes.

With this in mind, we now focus on probabilistic metrology. 
As discussed in the introduction, we can improve performance if we give up on the idea of deterministic protocols, by allowing for failures to perform the tasks they have been designed for.  %Probabilistic protocols may sometimes fail to perform the task for which they have been designed, in which case a warning is flagged. 
Accordingly, probabilistic metrology protocols
%A probabilistic protocol performs a given task with some probability of success. 
%As discussed in the introduction, by giving up on the idea of performing a task deterministically one may be able to design better protocols for some tasks. the there are multiple examples of probabilistic protocols that perform better in certain tasks, if one gives up on the idea on performing the task deterministically, i.e., by allowing for a certain probability of failure.
%As we show here metrology is no exception. 
will either succeed and provide a precise estimate $\hat\theta$, or warn of failure (abstain). 
Following these premises, the figure of  merit for such protocols are given~by the average uncertainty of the successful outcomes, i.e.,~by
\begin{equation}
\label{Loss-prob} 
L_{\rm s} = \int\! {\rm d} \hat\theta\, p(\hat\theta | \mathrm{succ}) L_{\hat\theta} .
\end{equation}
Conditional expectations such as this are the cornerstone of bayesian estimation. Their use is wide-spread and established in a number of disciplines, such as control theory or signal processing, where an accurate and rigorous assessment of the precision is required ---see for instance \cite{condest}.
In order to give a complete characterization of the probabilistic protocol, one should supplement the attained uncertainty $L_{\rm s}$, with the corresponding probability of success,~$S$. We will derive the tradeoff curve $L_{\rm s}(S)$ that gives the minimum uncertainty for every fixed value of the success probability~$S$.  In particular, by computing $\lim_{S\to0}L_{\rm s}(S)$ we will show that there is an \emph{ultimate quantum limit}  in the precision of an estimate inferred from {\em any outcome} of a quantum measurement. %Returning to the hypothetical metrology experiment above. For some fixed resources the precision given by the deterministic bounds may not meet the requirements of the experimenter (e.g. to publish the result it in some book of standards). Nonetheless, the use of a probabilistic scheme might well be able to offer the required precision upon a heralded outcome.

At this point, it should be clear that a probabilisitic protocol, as defined above, is not meant to be repeated %a given number of times till 
until it succeeds~\cite{combes_quantum_2014}. Obviously, such a strategy would be ultimately deterministic (it will always end up providing an estimate) and, thus, it could not outperform the optimal deterministic protocol for the same total amount of resources. %The goal of probabilistic metrology is to provide a scheme that, with some pre-established success probability provides a guaranteed precision.
Only with some pre-establish success probability can probabilistic metrology provide a guaranteed precision for a given amount of resources.% with some pre-established success probability.

We next outline an alternative approach that is often used in quantum metrology and point out the differences with the global single-shot framework defined above.
%
%Before finishing this section, we outline the 
%We finish this section by comparing the global single-shot framework defined above to one that seems more inspired by the frequentist approach that is also widely used in the quantum metrology community (pointwise and asymptotically large number of repetitions). 
The so-called pointwise approach aims at minimizing the dispersion of the estimates~$\hat\theta$ that results from the noise inherent to quantum measurements. It assumes that the true value of the parameter $\theta$ is fixed (i.e., it is not a random variable), and that the metrology protocol can be repeated an arbitrary number of times; it is a frequentist framework.
%The frequentist framework has the notion of ensemble deeply rooted in it and it implicitly assumes the experiment can be repeated an indefinite number of times.  
%A natural and very common figure of merit is the mean square error 
It is customary to quantify the precision of the protocol by the mean square error,
\begin{equation}
\mathrm{MSE}_{\theta}=%L_{\theta}=
\int\! {\rm d} \hat\theta\, p(\hat\theta | \theta)\, \ml_{\rm q}(\theta,\hat{\theta}),
\end{equation}
which indeed gives a measure of how the estimates would scatter around the true value if the protocol is repeated many times. Note that if a prior $\pi(\theta)$ were supplied, one could compute the average over $\theta$, thus recovering the expression of the Bayesian expected loss in Eq.~\eqref{Loss-expected} for the quadratic loss function. 

The celebrated quantum Cramer-Rao bound~\cite{braunstein_statistical_1994} provides a lower bound to ${\rm MSE}_{\theta}$ that can be readily computed. In addition, one can often argue that the 
Cramer-Rao bound can be attained in the asymptotic limit of large number of resources by a suitable two-step adaptive protocol. However, the assumptions under which the quantum Cramer-Rao bound holds, and the conditions under which the bound is attained entail some subtleties that  are often ignored and that can lead to erroneous conclusions  \cite{berry_optimal_2012,giovannetti_sub-heisenberg_2012} %wrong bounds, 
 and misleading accounting of resources % even in the asymptotic regime of many such resources 
(see for instance \cite{hayashi_comparison_2011, hall_2012}). In the particular case of probabilistic metrology, the direct application of the pointwise approach can lead to unphysical results, as pointed out in~\cite{chiribella_quantum_2013}. 

The Bayesian approach has been widely used to assess the performance of quantum information protocols such as teleportation, state estimation, universal cloning and quantum memories. Despite its many advantages, which include a straightforward accounting of resources and its validity %for large as well as for a small number of resources 
even for a small (non-asymptotic) number of resources, it also has some drawbacks: optimal bounds are usually hard to compute and there is no general prescription to choose the  prior $\pi(\theta)$. 
In the case at hand (estimation of a phase $\theta$) these drawbacks can be easily evaded, as the symmetry of the problem simplifies the calculations significantly while providing a valid justification (Laplace’s principle of insufficient reason) to choose a uniform prior on $(-\pi$, $\pi]$.

%Its main
%advantages are: A1) Transparent. The  hypothesis are explicitly stated and easy to grasp; A2) The allocation of resources is straight-forward, A3) the results are valid both for asymptotically large and finite amount of resources. It's main disadvantages are: D1) Hard to compute optimal bounds. D2) no general prescription on how to fix a prior $\pi(\theta)$. In the case of phase estimation under study both can be  tackled. The symmetry of this problem a simply the calculations significantly and in addition it provides a valid justification (Laplace’s principle of insufficient reason) to fix a uniform prior on the unknown parameter between $-\pi$ and $\pi$.  

There is still another approach that suits our framework and does not require a prior distribution: the mini-max approach, whereby the average over the unknown parameter $\theta$ is replaced by its worst-case value:
\begin{equation}
\label{Losswc} 
L_{\mathrm{wc}}   =  \sup_{\theta\in (-\pi,\pi] }  \int  {\rm d} \hat \theta  \,    p(\hat \theta | \theta , \mathrm{succ})\,  \ml(\theta,\hat\theta),
\end{equation} 
where the optimization is over all possible quantum measurements. As shown in~\ref{sect:wc=bay}, for phase estimation the optimal worst-case loss, Eq.~(\ref{Losswc}), and the expected loss, Eq.~(\ref{Loss-prob}) are equivalent.

%--TOTAL amoun of resources. Claim optimal bound via CR should include an assesment of how $\nu$ (rep) scales with N; in particular $k$ should not increas with N up to inf (Pezzé PRL 100 073601):
%the number of measurements needed to achieve the asymptotic behavior predicted by the CRLB might depend on the mean number of particles [an example was discussed in A. S. Lane, S. L. Braunstein, and C. M. Caves, Phys. Rev. A 47, 1667 (1993)

\section{Results \label{sec:results}}

\begin{figure}[htbp] %  figure placement: here, top, bottom, or page
   \centering
   \includegraphics[width=3.6in]{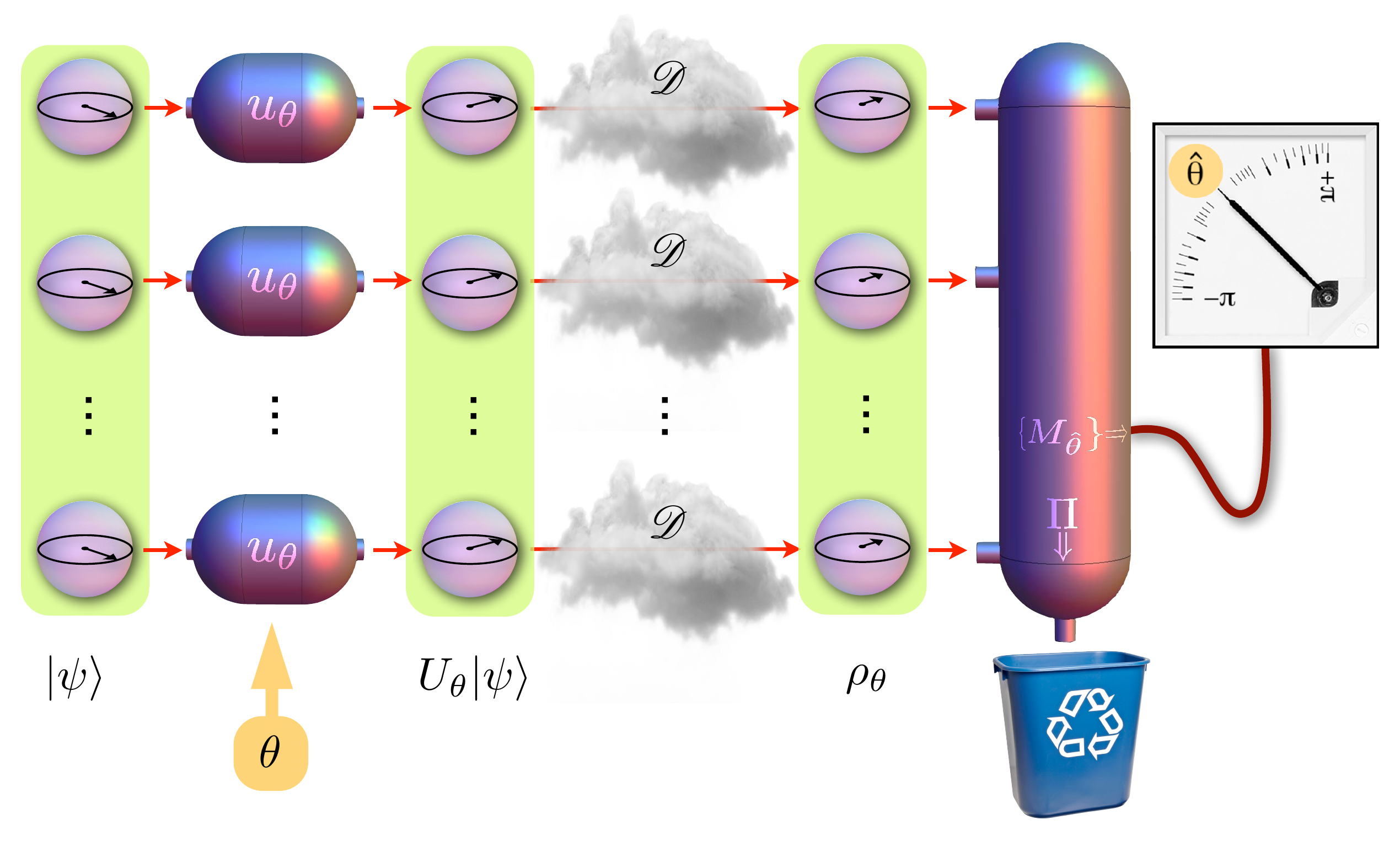} 
   \caption{{\bf Probabilistic Metrology protocol.} Pictorial representation of a probabilistic metrology protocol with $n$ qubits (depicted by small Bloch spheres). The probe state $|\psi\rangle$, which needs not necessarily be a product of identical copies, undergoes an evolution $U_\theta=u^{\otimes n}_\theta$ controlled by the unknown parameter $\theta$. Experimental noise $\mathscr D$ decoheres the system  before a collective measurement on \emph{all} qubits is performed. The measurement apparatus  either returns an ultra-precise estimate~$\hat{\theta}$ of the parameter or shows a failure signal. In the event of a failure, some information could be in principle scavenged (see last subsection in Results).}
   \label{fig:scheme}
\end{figure}

\subsection{Optimal probabilistic measurement for $n$-qubits \label{subs:nqubits}} 
In the scope of this paper, metrology aims at estimating the parameter $\theta$ that determines the unitary evolution, $U_\theta:=u^{\otimes n}_\theta$, of a probe system of $n$ qubits in the presence of local decoherence, where $ u_\theta=\exp (i \theta\ketbrad{1}) $.%, $N$ is the number operator.

As depicted in Fig.~\ref{fig:scheme}, the initial $n$-partite pure state~$|\psi\rangle\langle\psi|=\psi$ (this shorthand notation will be used throughout the paper) is prepared and is let evolve. The state is affected by uncorrelated dephasing noise, which can be modeled by independent  phase-flip errors occurring  with probability $p_f=(1-r)/2$ for each qubit.
Its action on the $n$-qubits is described by a map ${\mathscr D}$ that commutes with the Hamiltonian, so that it could as well be understood as acting before or during the phase imprinting process. 

Next, the experimentalist performs a suitable measurement on $\rho_\theta= {\mathscr D}(U_\theta\psi U^\dagger_\theta)$ and, based on its outcome, decides whether to abstain or to produce an estimate~$\hat\theta$ for the unknown parameter~$\theta$. Note that this decision is based solely on the outcome of the measurement as,  naturally, the actual value of $\theta$ is unknown to the experimentalist. Our aim is to find the optimal protocol, e.g.,  the measurement that gives the most accurate estimates for a given probe state and for a given maximum probability of abstention.

Motivated by the periodicity of the phase, we quantify the uncertainty of the estimated phase~$\hat\theta$ by the periodic loss function $\ml_{\rm p}(\theta,\hat\theta)$ defined in Sec.~\ref{sec:framework}, and to assess the performance of the protocol we use the expected loss defined in Eq.~\eqref{Loss-prob} and~Eq.~\eqref{Loss-theta}.
\begin{eqnarray}
\label{fidelity-def} 
\sigma^{2}\equiv L_{\rm s}&=& \int\!\!\int\! {\rm d} \hat\theta\,{\rm d}\theta   \,    p(\theta | \hat{\theta})p(\hat\theta | \mathrm{succ})\,  \ml_{\rm p}(\theta,\hat\theta)\nonumber \\
&=&\int\! \!\int\! {\rm d} \hat\theta\,{\rm d}\theta\, p(\theta,\hat\theta | \mathrm{succ}) \ml_{\rm p}(\theta,\hat\theta)\nonumber\\
&=&\frac{1}{S}\int\!\! \int\! {\rm d} \hat\theta\,{\rm d}\theta\, p(\theta,\hat\theta, \mathrm{succ}) \ml_{\rm p}(\theta,\hat\theta) ,
\end{eqnarray} 
where the success probability is $S= \int\! \!\int\! {\rm d} \hat\theta\,{\rm d}\theta\, p(\theta,\hat\theta, \mathrm{succ})$. Throughout the rest of the paper we will refer to $\sigma^2$ as the uncertainty for brevity. The uncertainty and the probability of success~$S$  will fully characterize our probabilistic metrology strategies. %The choice of the periodic loss function is motivated by the periodicity of the phase. 
In adition, in the asymptotic limit of large number of resources the distribution  $p(\hat\theta|\theta,\mathrm{succ})$  becomes peaked around the true value~$\theta$ and the uncertainty (expected loss)  approximates the mean-square error (expected loss for quatratic loss function $\ml_{q}$). 

The set $\{\rho_\theta\}$ %, that encode the unknown parameter $\theta$ 
is a so-called {\em covariant family of states}~\cite{holevo_probabilistic_1982}, as it is generated by the action of a group of unitaries; $\{U_\theta\}_{\theta\in (-\pi,\pi]}$ in our case. We also note that $ \ml_{\rm p}(\theta,\hat\theta)$ is invariant under the same group action, namely, $ \ml_{\rm p}(\theta+\theta',\hat\theta+\theta')= \ml_{\rm p}(\theta,\hat\theta)$ for all $\theta'\in(-\pi,\pi]$.
Because of this, there is no loss of generality %of the figure of merit in~\eqref{fidelity-def} and the covariance of $\{\rho_\theta\}_{\theta\in(-\pi,\pi]}$, one can show that there is no loss of generality 
in choosing the measurement to be covariant~\cite{holevo_probabilistic_1982}. Such {\em covariant measurements} are defined by~$\{M_{\hat\theta}=U_{\hat\theta}\Omega U^\dagger_{\hat\theta}/(2\pi)\}_{\hat\theta\in(-\pi,\pi]}$, where~$\Omega$
%, such that \mbox{$0\le\Omega\le\openone$}, 
is the so-called {\em seed} of the measurement. 
%Each operator is labeled by its corresponding estimate~$\hat{\theta}$, and
In addition, we have the invariant measurement operator \mbox{$\Pi=\openone- \int_{-\pi}^{\pi}d\hat\theta/(2\pi) U_{\hat\theta}\Omega\, U^\dagger_{\hat\theta}\le\openone$} that corresponds to the abstention event. 
With this, finding the optimal estimation scheme reduces to finding the operator $\Omega$  that mimimizes the uncertainty, 
\begin{eqnarray}
\sigma^{2}({S})&=&{1\over S}\min_{\Omega}\int_{-\pi}^\pi{d\hat{\theta}\over2\pi} \ml_{\rm p}(0,\hat\theta)\,{\rm tr}\!\left(U_{\hat\theta}\Omega U^\dagger_{\hat\theta}\rho\right),% \nonumber\\[.5em]
%&=&\frac{1}{2}\left(1+\frac{1}{S} \max_{\Omega} \sum_{b,b'} \Omega_{b,b'} \rho_{b',b}\delta_{|b'|,|b|+1}\right)
\label{F_max}
\end{eqnarray}
for a fixed success probability~\cite{gendra_quantum_2013}
\begin{equation}
S=\int_{-\pi}^\pi{d\hat \theta\over2\pi}{\rm tr}\!\left(U_{\hat\theta}\Omega U^\dagger_{\hat\theta}\rho\right) .%=
% \sum_{b,b'} \Omega_{b,b'} \rho_{b',b}\delta_{|b'|,|b|}.
\label{succ}
\end{equation}
In deriving Eq.~(\ref{F_max}) we have used covariance to fix the value of $\theta$ to zero and thereby  get rid of the integral over~$\theta$ in Eq.~\eqref{fidelity-def}, and have defined $\rho=\mathscr D(\psi)$ accordingly.

\medskip

\subsection{Symmetric probes.} 
We now focus on probe states consisting $n$-qubits that are initially prepared in a permutation invariant state. This family includes most of the states considered in the literature, our case-study of multiple copies of equatorial-states, and also, as we will show below, the optimal probe-state for probabilistic metrology. The input state is given 
by,
% We start our analysis by studying the power of abstention in the ubiquitous case of a probe state consisting of $n$ identical copies of equatorial qubits: 
\begin{equation}
\label{eq:symm}
\ket{\psi}=
\sum_{m=-J}^{J} c_{m}\ket{J,m},
\end{equation}
where $J=n/2$ is the maximum total spin angular momentum (hereafter spin for short) of~$n$ qubits %(with no loss of generality, we think of them as spin $1/2$ particles) 
and the set of states \mbox{$\{\ket{J,m}\}_{m=-J}^J$} spans the fully-symmetric subspace.   Given the permutation invariance of the noisy channel, the state $\rho=\mathscr{D}(\psi)$ inherits the symmetry of the probe, and can be conveniently written in a block diagonal form in the total spin bases~\cite{cirac_optimal_1999,bagan_optimal_2006} (see~\ref{subs:symmetricprobes}),  
 \begin{equation}\label{eq:rhoj}
 \rho=\sum_{j} p_{j} \rho^{j}\otimes \frac{\id_{j}}{\nu_{j}}, % \quad  \nu_{j} =\binom{2J}{J-j} \frac{2j +1}{J+j+1},
 \end{equation}
where the state $\rho^{j}$ has unit trace, $p_j$ is the probability of $\rho$ having spin~$j$, and $\id_{j}$ stands for the identity in the $\nu_j$-dimensional multiplicity space of the irreducible representation of  spin $j$. The  sum over $j$ in Eq.~\eqref{eq:rhoj} runs from
 $j_\mathrm{min}=0$ ($j_\mathrm{min}=1/2$) for  $n$ even (odd) to the maximum  spin~$J$.   Similarly, the measurement operators, can be taken to have the same symmetry and thus be of the form $\Omega=\sum_{j}\ketbrad{\chi_{j}}\otimes \id_{j}$, where $\ket{\chi_{j}}=\sum_{m}f^{j}_{m}\ket{j,m}$, $0 \leq f^{j}_{m}\leq 1$.
The minimum uncertainty $\sigma^2(S)$ for a fixed probability of success $S$ can hence be expressed in terms of the uncertainty $\sigma^{2}_j(s_j)$  in each irreducible block and its corresponding success probability $s_{j}$,
\begin{equation}
\label{eq:Fs}
\sigma^2(S)= \min_{\{s_{j}\}}  \sum_{j} \frac{p_{j} s_{j}}{S} \sigma^{2}_j(s_{j}),\quad S=\sum_{j}p_{j}s_{j},
\end{equation}
%
%such that $S=\sum_{j}p_{j}s_{j}$, 
where $ \sigma^{2}_j(s_{j})$ [$s_j$] is defined by Eq.~(\ref{F_max}) [Eq.~(\ref{succ})] with %$\Omega\to\chi_j$ 
$\Omega$, $\rho$ and $U_{\hat\theta}$ projected onto the subspace of total angular momentum $j$. Recalling that $U_{\hat\theta}=\sum_j U^{j}_{\hat\theta}\otimes \openone_j$, $[U^{j}_{\hat\theta}]_{m,m'}=\delta_{m,m'}{\rm e}^{i m \hat\theta}$ ($-j\le m,m'\le j$) and $\ml_{\rm p}(0,\hat\theta)=2-{\rm e}^{i\hat\theta}-{\rm e}^{-i\hat\theta}$, one can easily integrate $\hat\theta$ to obtain
\begin{eqnarray}
&\displaystyle \sigma^{2}_j(s_{j})=2-\frac{2}{s_{j}} \max_{0 \leq f^{j}_{m}\leq 1} \sum_{m}\! f^{j}_{m} \rho^{j}_{m,m+1}  f^{j}_{m+1},&\nonumber \\[.5em]
&\kern3em\displaystyle\mbox{subject to  }s_{j}=\sum_{m }(f^{j}_{m})^{2} \rho^{j}_{m,m}.&
\label{eq:fidj}
\end{eqnarray}

This formulation of the problem allows for a natural interpretation of the probabilistic protocol as a two step process: i) a stochastic filtering channel
\begin{equation}
\label{eq:stoch}
{\mathscr F}(\rho)=\Phi\, \rho\, \Phi,\qquad \Phi=\sum_{j,m}f^{j}_{m} \ketbrad{j,m}\otimes \id_{j},
\end{equation}
that  coherently  transforms each basis vector as \mbox{$\ket{j,m}\to f^{j}_{m}\ket{j,m}$}, so that it modulates the input to a state with enhanced phase-sensitivity, followed by ii)~a canonical covariant  measurement with seed \mbox{$\tilde\Omega=\sum_{j}\sum_{m,m'}\ket{j,m}\!\bra{j,m'}\otimes \id_{j}$} performed on the transformed state from which the value of the unknown phase is estimated.

By defining the vector $\xi^j$ with components given by $\xi^{j}_{m}=f_{m}^{j} (\rho^{j}_{m,m}/s_{j})^{1/2}$ and introducing the tridiagonal symmetric matrix $H^j$, with entries
\begin{eqnarray}
H^{j}_{m,m'}&=&2\delta_{m,m'}-a^{j}_{m}\delta_{m,m'-1}-a^j_{m'}\delta_{m-1,m'},\nonumber \\[.0em]
\kern1.5em a^{j}_{m}&=&\frac{\rho^{j}_{m,m+1}}{\sqrt{\rho^{j}_{m,m}\rho^{j}_{m+1,m+1}}},
\label{eq:H^j_mm}
\end{eqnarray}
%
%%
%\begin{eqnarray}
%a^{j}_{m}&=&\frac{\rho^{j}_{m,m+1}}{\sqrt{\rho^{j}_{m,m}\rho^{j}_{m+1,m+1}}}=\\
%&=&
%{\sqrt{\frac{j+m+1}{j+m}} \sum_{k} r^{2 {k}+1}\binomial{j+m}{{k}}\binomial{j-m}{{k}+1} \over
%\sqrt{\sum_{k} r^{2 {k}}\binomial{j+m}{{k}}\binomial{j-m}{{k}}
%\sum_{k'} r^{2 {k'}}\binomial{j+m+1}{{k'}}\binomial{j-m-1}{{k'}}}}\nonumber
%\end{eqnarray}
%%
 we can easily recast the former optimization problem as,
\begin{eqnarray}
\label{eq:fidjmax}
& \kern5em \msej :=\min_{\ket{\xi^{j}}} \bra{\xi^{j}}H^{j} \ket{\xi^{j}},  &\\[.5em]
 \label{eq:conds}
&\displaystyle \mbox{subject to  } \braket{\xi^{j}} {\xi^{j}}\!=\!1\  \mbox{and } 0\!\le\! \xi^{j}_{m}\!\leq\! (\rho^{j}_{m,m}/\!s_{j})^{1/2}\!.& %\nonumber
\end{eqnarray}
%%
%where $M^{j}_{m,m'}=a^{j}_{m}\delta_{m,m'-1}+a^j_{m'}\delta_{m-1,m'}$ with
%%
%\begin{eqnarray}
%a^{j}_{m}&=&\frac{\rho^{j}_{m,m+1}}{\sqrt{\rho^{j}_{m,m}\rho^{j}_{m+1,m+1}}}=\\
%&=&
%{\sqrt{\frac{j+m+1}{j+m}} \sum_{k} r^{2 {k}+1}\binomial{j+m}{{k}}\binomial{j-m}{{k}+1} \over
%\sqrt{\sum_{k} r^{2 {k}}\binomial{j+m}{{k}}\binomial{j-m}{{k}}
%\sum_{k'} r^{2 {k'}}\binomial{j+m+1}{{k'}}\binomial{j-m-1}{{k'}}}}\nonumber
%\end{eqnarray}
%%
Note that $a^j_m$, and in turn $H^j$, depend on the strength of the noise but they take the same values for {\em all} symmetric probe states, since $\rho^{j}_{m,m'}\propto c_{m}c_{m'}$. 
%It will be convenient to define the matrix $H^{j}:=2\id -M^{j}$, so that the conditional fidelity in  \eqref{eq:fidjmax}
%can be written as
%%
%\begin{equation}
%\label{eq:fidjH}
%F_j(s_j)=1-\frac{1}{4}\msej,\quad  \msej :=\min_{\ket{\xi^{j}}} \bra{\xi^{j}}H^{j} \ket{\xi^{j}},
%\end{equation}
%
%subject to the constrains \eqref{eq:conds}. 
%
For deterministic strategies ($S=1$, i.e., $s_{j}=1$ for all $j$) no minimization is required and one only needs to evaluate the expectation values of $H^{j}$ for the `state' $\xi^{j}_{m}=(\rho^{j}_{m,m})^{1/2}$.
For~large enough abstention, %, $S\leq S^*$, 
the problem becomes an unconstrained minimization, so  $\msej$  is the minimal eigenvalue of $H^{j}$, and $\ket{\xi^{j}}$  its corresponding eigenstate.   From Eq.~\eqref{eq:conds}, we find that the corresponding filtering operation only succeeds with a probability %Eq.~\eqref{eq:conds}
\begin{equation}\label{eq:Scrit}
S^{*}=\sum_{j}p_{j}s_{j}^{*} ,\qquad  s_{j}^{*} =\min_{m}\frac{\rho^{j}_{m,m}}{{\xi^{j}_{m}}^{2}} .
\end{equation}	
We will refer to $S^{*}$ as the critical success probability, since the precision will not improve by decreasing the success probability below this value: $\sigma^2(S)=\sigma^{2}(S^{*})$ for $S\leq S^{*}$.

\medskip

\subsection{Asymptotic scaling: particle in a potential box}
 In order to compute the scaling of the uncertainty as the number of resources becomes very large we need to solve the above optimization problem in the asymptotic limit of~$n\to \infty$. We start be analyzing the uncertainty $\sigma^{2}_j(s_{j})$ for blocks of large~$j$.
As shown in~\ref{subs:continuum},  for each such block we define the ratios $x= m/j$, $m=-j,-j+1,\dots,j$, that approach a continuous variable  as~$j\to\infty$. In this limit, $\{\sqrt j \xi_m^{j}\}$ approaches a real function of $x$, $\sqrt j \xi_m^{j}\to\varphi(x)$, and the expectation value in Eq.~\eqref{eq:fidjmax} becomes,
\begin{eqnarray}
\label{eq: sigmaj2}
 \msej &=&\!{
1\over j^2}\min_{\ket{\varphi}}%
%{\varphi_j(-1)^{2}+\varphi_{j}(1)^{2}\over j}\right.\nonumber\\[.5em]
%&+&\left.
\int_{-1}^1 \!\!dx  \left\{\left[{d\varphi(x)\over dx}\right]^2\!\!+{V^j(x)}\varphi(x)^{2}\right\}\nonumber\\[.5em]
&:=&{
1\over j^2}\min_{\ket{\varphi}}\bra{\varphi} {\cal H}^j\ket{\varphi},
\end{eqnarray}
where we have dropped some boundary terms that are irrelevant for this discussion,  and where ${\cal H}^j:=-d^2/dx^2+V^j(x)$ plays the role of a `Hamiltonian', with a `potential'
\begin{equation}
V^j(x)= 2 j^2 (1-a^j_m)= j{1-r^2\over2r\sqrt{1-(1-r^2)x^2}} .
\label{eq:pot}
\end{equation}
Furthermore, in Eq.~(\ref{eq: sigmaj2}) the function $\varphi(x)$ must be also  differentiable and must satisfy the conditions
\begin{equation}
\label{eq:cond-cont}
\braket{\varphi} {\varphi}=\!\int_{-1}^1 \!\!\!dx\;[\varphi(x)]^{2}=1,\quad
\varphi(x)\leq \frac{\tilde\varphi(x)}{\sqrt{s_{j}}} ,
\end{equation}
%
%%
%\begin{eqnarray}
%\label{eq:cond-cont}
%&&\braket{\varphi_{j}} {\varphi_{j}}=\!\int_{-1}^1 \!\!\!dx\;\varphi_{j}\!(x)^{2}=1  \mbox{and } \\[.5em]
% &&\varphi_{j}(x)\leq \frac{\tilde\varphi_{j}(x)}{\sqrt{s_{j}}} \mbox{ with } \tilde\varphi_{j}(x=\frac{m}{j})=\lim_{j\to \infty} \sqrt{j \rho^{j}_{mm}}.\label{eq:kkt}
%\end{eqnarray}
%%
where for a given large $j$ we define $\tilde\varphi(x)$  through 
\begin{equation}
\sqrt{j \rho^{j}_{mm}} \to \tilde\varphi(x),
\quad x=\frac{m}{j}
.\label{eq:kkt}
\end{equation}
%
%Note that in order to keep the notation uncluttered we omit the subindex in $x_{j}$, here and throughout the rest of the paper. 
It is now apparent from Eqs.~(\ref{eq: sigmaj2}) through~(\ref{eq:kkt}) that our optimization problem is formally equivalent to~that of finding the ground-state wave-function of  a quantum particle in a box ($-1\le x\le1$) for the potential~$V^j(x)$ and subject to boundary conditions that are fixed by  the probe state,  the strength of the noise, and the success probability. Other equivalent variational formulations can be found in \cite{gendra_quantum_2013,gendra_optimal_2013, summy_phase_1990} for pure states and in \cite{knysh_true_2014} for the pointwise approach.

% \footnote{The analogy with the quantum mechanical problem of the
%  groundstate of a particle in a potential was first presented in: J. Calsamiglia,  Abstention-enhanced metrology, Noise Information \& Complexity @ Quantum Scale, Ettore Majorana Centre, Erice, Italy (2013). Other equivalent variational formulations can be found in \cite{gendra_quantum_2013,gendra_optimal_2013, summy_phase_1990} for pure states; and in the  in \cite{knysh_true_2014} for the pointwise approach.} -

\medskip

\subsection{Multiple-copies.} Although our methods apply to general symmetric probes, for the sake of concreteness we study in full detail the paradigmatic case of a probe consisting of $n$ identical copies of equatorial qubits:
\begin{equation}
\label{eq:multicopy}
\ket{\psi_{{\rm cop}}}= \frac{1}{\sqrt{2^{n}}}( \ket{0}+\ket{1})^{\otimes n}. %=
%\sum_{i=0}^n c_{m}\ket{J,m},
\end{equation}
%
%where 
%$c_{m}=2^{-n/2}{{n}\choose{n/2-m}}^{-1/2}$ and 
%iid stands for independent and identically distributed.  
Decoherence turns this symmetric pure state to a full rank state with a probability of having  spin $j$ given~by
 \begin{equation}\label{eq:pjmulticopy}
 p_{j}\simeq {{\rm e}^{-J{(j/J-r)^2\over 1-r^2}}\over\sqrt{\pi J(1-r^2)}},
 \end{equation}
where this approximation is valid around its peak, at the typical value \mbox{$j_{0}=r J$}. %and whose width reads \mbox{$\Delta{j}=[(1-r^{2}) J/2]^{1/2}$}.
For each irreducible block and before filtering we have a signal
\begin{equation}
\label{eq:gaussian}
\sqrt{j \rho^j_{mm}}\to\tilde\varphi(x)\simeq\left(\frac{j r}{\pi}\right)^{\frac{1}{4}}\expo{-\frac{ r j}{2} x^{2}}
\end{equation}
that peaks at $x=0$ with variance $\mean{x^{2}}=({2 r j})^{-1}$.

For deterministic protocols ($S=1$)  the constraints completely fix the solution: $\varphi(x)=\tilde\varphi(x)$. The corresponding uncertainty is obtained by computing the `mean energy' \mbox{$\msej=\mean{{\cal H}^j}_{\tilde\varphi}/j^2$}, in Eq.~\eqref{eq: sigmaj2}. For large $j$ it is meaningful to use the harmonic approximation $V^j(x)\simeq V^j_0+\omega_{j}^{2} x^2$, where $V^j_0=j(1-r^2)/(2r)$ and $\omega_{j}^{2}=j(1-r^2)^2/(4r)$. The leading contribution to $\msej$ comes from the `kinetic energy' [i.e., the first term in Eq.(\ref{eq: sigmaj2})], which gives $\mean{p^{2}}_{\tilde\varphi}=(1/4) \mean{x^{2}}^{-1}=j r/2$, whereas the harmonic term gives a sub-leading contribution. One easily obtains $\msej=(2 j r)^{-1}$. The leading contribution to the uncertainty of the deterministic protocol is given by $\msej$ at the typical  spin~$j_0$: $\sigma^2_{\rm det}=(2Jr^2)^{-1}=(nr^2)^{-1}$, in agreement with the previous known (pointwise) bounds.

%
%For this purpose we expand the potential around $x=0$:
%$V(x)=j \frac{1-r^{2}}{2 r}+ j \frac{(1-r^{2})^{2}}{4 r}x^{2}$. The first constant term is of the same order than the `kinetic' energy ($\mean{p^{2}}=1/2 \mean{x^{2}}^{-1}=j r$) while the harmonic contribution is of a sub-leading order, hence $\msej=\frac{1}{2 j r}$. Since the probability $p_{j}$ in \eqref{eq:pjmulticopy} is strongly peaked at $j_{0}=nr/2$, the asymptotic expression for the precision can readily be written as $\mse_{\mathrm{det}}=\frac{1}{n r^{2}}$, which agrees with the known bounds \cite{}[bagan? D'ariano].

For unlimited abstention in a block of given spin~$j$ ($s_{j}$ very small) the minimization in Eq.~(\ref{eq: sigmaj2}) is effectively unconstrained and the solution 
(the filtered state) is given by the ground state $\varphi^{\rm g}(x)$ of the potential $V^j(x)$.
Within the harmonic approximation,  we notice that the effective frequency of the oscillator grows as $\sqrt{j}$, and the corresponding gaussian ground state is confined around~$x=0$ with variance $\mean{ x^{2}}=(r/j)^{1/2}(1-r^2)^{-1}$. In this situation both the kinetic and harmonic  contributions to the `energy' are sub-leading, and so are the higher order corrections to $V^j(x)$. Thus, the uncertainty $\msej$ for  spin~$j$ is ultimately limited by the constant term~$V^j_0$ of the potential. Up to sub-leading order one obtains 
  $\msej=(1-r^2)(2jr)^{-1}[1+(r/j)^{1/2}]$. The filtering of~$\tilde\varphi(x)$ to produce the  gaussian ground state $\varphi^{\rm g}(x)$ succeeds with probability  $s^{*}_{j}\sim \expo{-2 j \log(1+r)}$ (see~\ref{subs:relevantExpres}). Note that in the absence of noise ($r=0$) the potential $V^j(x)$ vanishes and the ground state is solely confined by the bounding box~$-1\le x\le1$. Then,  $\varphi^{\rm g}(x)=\cos(\pi x/2)$, which results in a Heisenberg limited precision (the ultimate pure-state bound):~$\mse=\pi^2/n^2$~\cite{summy_phase_1990,gendra_quantum_2013}.

 If the optimal filtering is performed on typical blocks,  $j \approx j_0$, one obtains $\mse=(1-r^{2})/(n r^{2})$, which  coincides with the ultimate deterministic bound found in~\cite{demkowicz-dobrzanski_elusive_2012,knysh_true_2014}.  
This shows that a probabilistic protocol that uses the uncorrelated multi-copy probe state $|\psi_{\rm cop}\rangle$ can attain the  
 precision bound of a deterministic protocol, for which a highly entangled probe is required. This bound is attained for a critical success probability \mbox{$S^{*}\simeq s^*_{j_{0}}\sim  \expo{-n r \log(1+r)}$}.
 More interestingly, we can push the limit further by post-selecting on the block with highest spin (by choosing~$f^{j}_{m}\propto \delta_{j,J}$) to obtain
 \begin{equation}
 \label{eq:ultimate}
\mse_{\mathrm{ult}}%\equiv \mse(S\to 0)=
:=\mse_{j\approx J}=\frac{1-r^{2}}{n r}  \left(1+\sqrt{\frac{2 r}{n}}\right),
  \end{equation}
  %
 %where we have included the sub-leading correction. 
 with a critical probability given by $S^{*}= p_{J} s^{*}_{J}\sim\expo{-n  \log 2}$, independently of the noise strength. 
 We note that the leading order is a factor $r$ smaller than the previously established (deterministic) bound, $\mse=(1-r^{2})/(n r^{2})$\cite{demkowicz-dobrzanski_elusive_2012,knysh_true_2014}. This important enhancement in precision results from post-selection of high-angular momentum, which does not commute with the noise channel. Hence, in contrast to the noiseless scenario, post-selection is not equivalent to a suitable choice of input state.
 
Having understood the two limiting cases of no abstention (the deterministic protocol) and unlimited abstention, we can now quantify the asymptotic scaling for an arbitrary success probability $\mse({S})$. We use the Karush-Kuhn-Tucker optimization method to minimize Eq.~(\ref{eq: sigmaj2}) under the constraints in Eq.~\eqref{eq:cond-cont}. For a given value of $s_{j}$, the so-called complementary slackness condition~\cite{boyd_convex_2004,gendra_optimal_2013} guarantees that 
the solution~$\varphi(x)$ to Eq.~(\ref{eq: sigmaj2}) saturates the inequality in Eq.~\eqref{eq:cond-cont} for $x$ in a certain region called coincidence set, while it coincides with an eigenfunction of the Hamiltonian~${\cal H}^j$, defined after Eq.~(\ref{eq: sigmaj2}), for $x$ outside this region. The continuity of~$\varphi(x)$ and its derivative provide some matching conditions at the border of the coincidence set and a unique solution can be easily found. 

\begin{figure}[htbp] %  figure placement: here, top, bottom, or page
  \centering
   \includegraphics[width=3.4in]{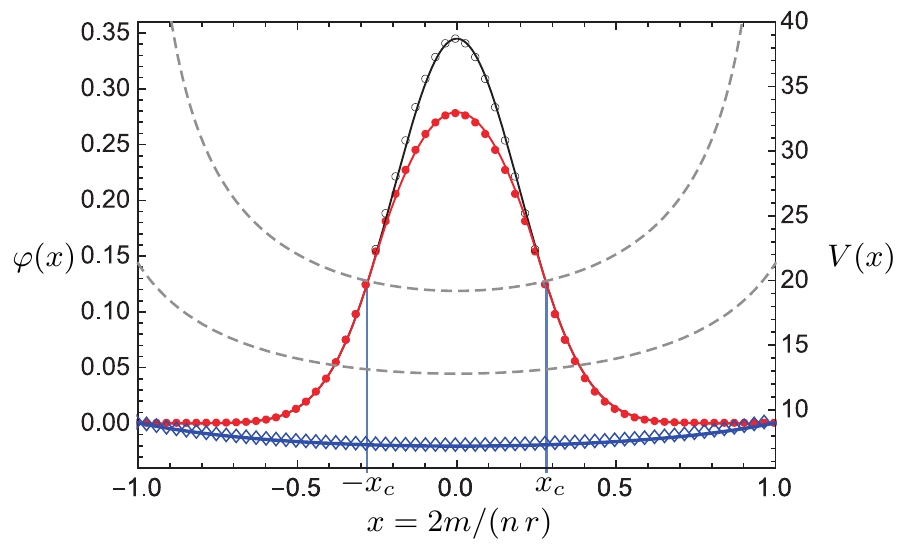} 
   \caption{{\bf Potential box equivalence}: Computing the action of the probabilistic filter and its precision is formally equivalent to 
   computing the ground state and energy of a particle in a one-dimensional potential box. The  state $\tilde\varphi(x)$  (empty circles) before the probabilistic filter and the state $\varphi(x)$ (solid circles) after the filter are represented together with the  potential $V(x)$ (diamonds), corresponding to $j=n r/2$ [see Eq.~\eqref{eq:pot}], for success probability $S=0.75$, noise strength $r=0.8$ and $n=80$ probe copies. The unfiltered state (empty circles) has been rescaled so that it coincides with the filtered state in the region $|x| \geq x_{c} = 9/32$. The effective potential depends on the noise strength, as illustrated by the two additional dashed curves: for $r=0.2$ (above) and $r=0.6$ (below). Numerical (symbols) and analytical results (lines) are in full agreement.}
   \label{fig:profile}
\end{figure}

\begin{figure}[htbp] %  figure placement: here, top, bottom, or page
   \centering
   \includegraphics[width=3.4in]{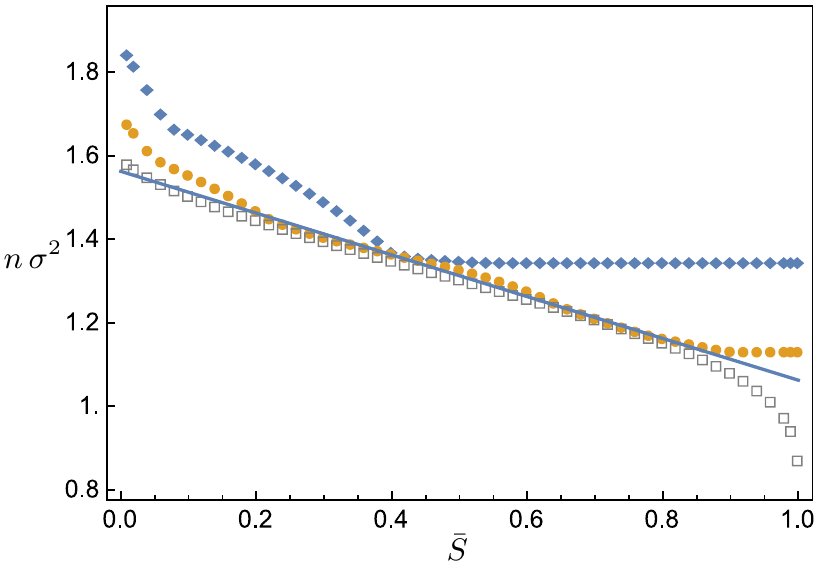} 
   \caption{{\bf Small \boldmath{$n$} precision.} 
Numerical results for the rescaled uncertainty~$n \sigma^2$ are plotted as a function of $\bar S=1-S$  for a noise strength of~$r=0.8$ and various number of copies: $n =6,10,20$ (diamonds, circles, squares).  
 The critical success probability, $S^*=1-\bar S^*$, is clearly identified at $\bar S^*\approx 0.46$  for~$n=6$, and at $\bar S^*\approx 0.9$ for $n=10$. For $n=20$, one has $S^*\!\approx1\!-5 \cdot\! 10^{-5}$, which is not visible in the figure.  The solid line is the analytical approximation on the right of Eq.~\eqref{eq:finiteQ}. }
   \label{fig:sigmaVSQ}
\end{figure}

As shown in Figure 2, in the case of multiple copies the tails of $\varphi(x)$ coincide with the gaussian profile in Eq.~(\ref{eq:gaussian}) scaled by the factor $s_j^{-1/2}$ for $|x|>x_{c}$ (in the coincidence set), while for $|x|<x_{c}$ the filter takes an active part in reshaping the peak into the optimal profile. Clearly, the wider the filtered region, the higher the precision and the abstention rate. 
A simple expression for the leading order can be obtained if we notice that with a finite abstention probability one can change the variance of the wave function in Eq.~\eqref{eq:gaussian} but not its $1/j$ scaling. Hence, as for the deterministic case, only the kinetic energy and the constant term $V^j_0$ of the potential play a significant role. The solution can then be easily written in terms of the pure-state solution~\cite{gendra_optimal_2013}, which corresponds to a zero potential inside the box \mbox{$-1\le x\le1$}:
\begin{equation}
\label{eq:finiteQ}
\mse\simeq \mse_{j_{0}}=\frac{1-r^{2}}{n r^{2}}+r \mse_{\rm pure}(S)\approx\frac{1-(r^2/2) \bar S}{n r^{2}},
\end{equation}
where $\bar S:=1-S$ is the probability of abstention, $\mse_{\rm pure}$ is the uncertainty for pure states ($r=1$) and for an effective number of qubits $n_{\rm eff}=2j_0$. The pre-factor~$r$ takes into account the scaling of the variance of the state Eq.~\eqref{eq:gaussian} as compared to the pure-state case.
The first equality of Eq.~\eqref{eq:finiteQ} uses the fact that only abstention on blocks  about the typical spin~$j_{0}$ is affordable for finite $S$. This also fixes the value of~$S$ to be approximately~$s_{j_{0}}$. The simple expression on the right of Eq.~\eqref{eq:finiteQ} is not an exact bound, but does provide a good approximation for moderate values of $\bar S$ (see Figure~\ref{fig:sigmaVSQ}).
We notice that for low levels of noise ($r\approx 1$) one can have a considerable gain in precision already  for finite abstention.

 \subsection{Finite $n$.} Up to this point, we have given analytical results for asymptotically large~$n$, the number of resources. In order to get exact values for finite $n$ we need to resort on numerical analysis. The main observation here is that  our optimization problem  can be cast as a semidefinite program: 
 \begin{equation}
 \label{eq:SDP}
 \mse=\min_{\Lambda\,:\,{\cal C}} \tr{H\Lambda},
 \end{equation}
 subject to a set of linear conditions on the matrix~$\Lambda$ given by
 {${\cal C}:=\{ \Lambda\geq 0,\, {\rm tr}\,\Lambda=1,\, $} $\Lambda^{j}_{mm}\leq p_{j} \rho^{j}_{m,m}/S\}$, where $\Lambda$ and $H$ have the block diagonal form $\Lambda=\oplus_j \Lambda^{j}$ and~$H=\oplus_j H^{j}$. Semidefinite programming problems, such as this, can be solved efficiently with arbitrary precision~\cite{boyd_convex_2004}.

Figure~\ref{fig:sigmaVSQ} shows representative results for moderate, experimentally relevant number of qubits $n$. We plot the uncertainty as a function of the abstention probability~$\bar S$ and noise strength $r=0.8$. We observe that for small values of $n$ the precision increases ($n \sigma^2$ decreases) quite rapidly until the critical value~$S^{*}$ is reached. Past this point the precision cannot be improved. For larger~$n$, the initial gain is less dramatic, but the critical  point (or plateau) is reached for higher abstention probabilities, hence allowing to reach a higher precision. We see that for moderately large $n$, abstention can easily provide $60\%$ improvement of the precision. When $n$ is large enough, e.g.,~$n=20$ (see the figure), there is a sharp improvement in precision as the success probability approaches the critical value. In the asymptotic limit, $n\!\to\! \infty$, it gives rise to a critical behavior that interpolates between the ultimate precision limit, Eq. \eqref{eq:ultimate}, and 
the precision for finite values of $\bar S$, Eq.~\eqref{eq:finiteQ}.

\begin{figure}[htbp] %  figure placement: here, top, bottom, or page
   \centering
   \includegraphics[width=3.4in]{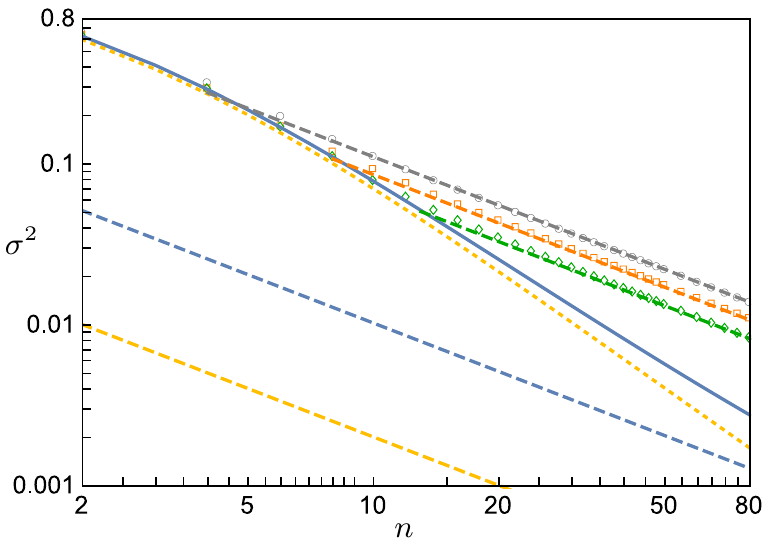} 
   \caption{{\bf \boldmath Moderate and large $n$ scaling.}  Ultimate precision scalings (for~$n\to \infty$) are of fundamental interest. However,  from the practical perspective, understanding the transient behavior is equally important. The plot shows the uncertainty $\mse$ as a function of $n$.  The circles, squares and diamonds are numerical results (SDP) for $r=0.95$, corresponding respectively to~$\bar S=0,\ 0.5,\ 0.9$. They fall on top of the (gray, orange, green) dashed straight lines given in Eq.~(\ref{eq:finiteQ}). 
 The solid blue line corresponds to the ultimate limit ($\bar S$ arbitrarily close to unity) for the same value of $r$, obtained via exact diagonlization of $H_{j}$ in \eqref{eq:fidjmax}. Its asymptotic expression, given by the leading order in Eq.~(\ref{eq:ultimate}), is the straight line plotted in dashed blue.  The ultimate limit for lower level of noise, $r=0.99$, is also plotted: \; the yellow dotted [dashed] line corresponds to the exact ultimate limit\;  [its asymptotic leading-order expression in Eq.~(\ref{eq:ultimate})].}
 \label{fig:sigmaVSQ2}
\end{figure}

Figure~\ref{fig:sigmaVSQ2} shows the scaling of the uncertainty with the amount of resources, $n$,  for low levels of noise $r=95\%$ and for different values of the abstention probability~$\bar S$. For~low~$n$ all curves exhibit a similar ($n^{-1}$  scaling, i.e., SQL). As we increase $n$, very soon  the curve corresponding to unlimited abstention (solid line) shows a big drop with a quantum-enhanced transient scaling given by $n^{-(\alpha+1)}$, where $\alpha>0$ depends on the noise strength. For very large $n$ ($\sim 500$) this curve saturates the ultimate asymptotic limit in Eq.~\eqref{eq:ultimate} (blue dashed line), which has again SQL scaling. The numerical results for finite $S$ (circles, squares, diamonds) display the optimal scaling up to the point where they meet the asymptotic (dashed) straight lines given by Eq.~\eqref{eq:finiteQ}. Past this point, they fall on top of the corresponding straight lines, which display SQL scaling.
The larger the abstention probability, the later this transition takes place. In addition,  the figure shows the ultimate scaling for $r=99\%$ to illustrate the fact that for weaker noise levels the transient is more abrupt ($\alpha $ is larger).

\medskip

\subsection{Ultimate bound for metrology}
So far we have studied the best precision bounds that can be attained for a fixed input state.
A very relevant question of fundamental and practical interest is whether this bounds can be overcome by an appropriate choice of  such state. We answer this question in the negative: the precision bound given by the uncertainty in Eq.~\eqref{eq:ultimate} is indeed the ultimate bound for metrology~in the presence of local decoherence and can only be attained by a probabilistic strategy.

To this aim, we first show in~\ref{subs:localdephasing} that for any probe state and any measurement that attain certain precision  (or, equivalently $\sigma^2$) with success probability~$S$, we can find  a new probe lying in the fully symmetric subspace ($j=J$) and a permutation invariant measurement that attain the very same precision with the very same success probability. This shows that the formulation that we have introduced,  with probe states in the fully symmetric subspace, is actually completely general. 

We now recall that the Hamiltonian  ${\cal H}^j$ is independent of the choice of probe state and that such choice determines only the shape of the state~$\tilde\varphi(x)$ before filtering, and the probability $p_{j}$ of belonging to the subspace of  spin $j$. Since the
bound in Eq.~\eqref{eq:ultimate} is attained by the ground-state $\varphi^{\rm g}(x)$ of the potential $V^J(x)$, the choice of probe cannot further improve the precision, but only change the success probability.
In particular one might increase $S$ by choosing a probe state that gives rise to a profile $\tilde\varphi(x)=\varphi^{g}(x)$ for $j=J$, without any  filtering within the block. In this case the critical success probability becomes $S^*=p_{J}={\rm e}^{-n[\log 2-\log(1+r)]}$ (see~\ref{subs:ultimateBwithout}), which is larger than that attained by $|\psi_{\rm cop}\rangle$. 

At the other extreme, for deterministic strategies, the calculation of $\sigma^2_{\rm opt}(1)$ can be easily carried out by performing first the sum over $j$ and then optimizing over the $(n+1)$-dimensional probe state.  In the continuum limit (large $n$) such calculation  can again be cast as  a variational problem formally equivalent to that of finding the ground-state of particle in a box with the harmonic potential $V(y)= n r^{-2}({1-r^{2}})(1+ y^{2})$, $-1\le y=m/J\le 1$. % (see methods).  %$\sigma^2=\bra{\psi} {\cal H}' \ket{\psi}$, where ${\cal H}'=d^2/dy^2+V(y)$, $y=m/J$, in the continuum limit. We arriving to a variational problem formally equivalent to that of the ground-state of particle in a box with a potential $V(x)= n \frac{1-r^{2}}{r^{2}}(1+ x^{2})$ (see methods). 
The corresponding ground state wave function and its energy provide the optimal probe state and uncertainty respectively:
\begin{equation}
\label{eq:probe}
\psi_{\mathrm{op}}(y)=\left[{n(1-r^2)\over(2\pi r)^{2}}\right]^{\frac{1}{8}}\expo{-{\sqrt{n(1-r^2)}\over4r}y^2} %;\quad x=2m/n
\end{equation}
and
\begin{equation}
\mse_{\mathrm{op}}(1)={1-r^2\over n r^2}+{2\sqrt{1-r^2}\over n^{3/2} r}.
\label{eq:opten}
\end{equation}
These results agree with their pointwise counterparts in~\cite{demkowicz-dobrzanski_elusive_2012,knysh_true_2014}. The presence of noise brings the pointwise and global approaches in agreement, as to both the attainable precision and the optimal probe state are concerned. This  agreement between global and pointwise approaches has been recently showed to be a generic feature in noisy scenarios with shot-noise limited precision~\cite{Jarzyna_2015}.
This is in stark contrast with the noiseless case, where the probe  $\psi(y)=\cos(y\pi/2)$ is optimal for the global approach and gives $\mse_{\rm opt}=\pi^{2}/n^{2}$, while
the NOON-type state $\ket{\psi}=2^{-1/2}(\ket{J,J}+\ket{J,-J})$ provides the optimal pointwise uncertainty~$\mse_{\rm opt}=1/n^{2}$. 

It remains an open question to find the optimal probe state given a finite values of $\bar S$.  As argued above, a finite $S$ will only be able to moderately reshape the profile without significantly changing the scaling of its width. Therefore, we expect the optimal state to be fairly independent of the precise (finite) value of $S$, and hence very close to that obtained for the deterministic case ($S=1$). Numerical evidence (optimizing simultaneously over probes and measurements) suggests that this is indeed the case provided $S$ is not too small. With this we are lead to conjecture that the optimal probe state is given by 
\begin{equation}
\label{eq:optFiniten}
c^{\mathrm{opt}}_{m}\propto\cos\left(\frac{m \pi}{n+2} \right) \expo{-{\sqrt{\frac{1-r^{2}}{ r^{2 } n^{3}}}} m^2},
\end{equation}
independently of $S$  (finite),  which  agrees with Eq.~\eqref{eq:probe}  for asymptotically large $n$. Note that the cosine prefactor guarantees that the solution converges to the optimal one for $r\to 1$ and  it keeps the state confined in the box for all values of $n$ and $r$. Such states continue to have a dominant typical value of $j=j_{0}$ and in those blocks both the kinetic and harmonic contributions to the energy are  of sub-leading order. Hence,  for  probes of the form in Eq.~\eqref{eq:optFiniten} the enhancement due to abstention is very limited, up until very high abstention probabilities where one can afford to post-select high spin states to reach the ultimate limit in Eq.~\eqref{eq:ultimate}.

\medskip

\subsection{Scavenging information from discarded events}
The aim of probabilistic metrology is twofold. First, it should estimate an unknown phase $\theta$ encoded in a quantum state with a precision that exceeds the bounds of the deterministic protocols. Second,  it should assess the risk of failing to provide an estimate at all (e.g., it should provide the probability of success/abstention). Probabilistic metrology protocols are hence characterized by a precision versus probability of success trade-off curve, or equivalently by~$\sigma^2(S)$. As such, no attention is payed to the information on $\theta$ that might be available after an unfavorable outcome.  
Here, we wish to point out that one can attain $\mse_{\rm opt}(S)$ and still be able to recover, or scavenge, a fairly good estimate from the discarded outcomes (see Fig.~\ref{fig:scheme}). 

The optimal scavenging protocol can be easily characterized in terms of the stochastic map~${\mathscr F}$ in Eq.~\eqref{eq:stoch}, which describes the state transformation after a favorable event, and that associated to the unfavorable events: 
\begin{equation}
{\bar {\mathscr F}}(\rho_{\theta})=\bar\Phi \rho_{\theta} \bar\Phi ,\quad \bar\Phi=\sum_{j,m}\bar{f}^{j}_m \ketbrad{j,m}\otimes \id_{j} ,
\end{equation}
where the weights $\bar{f}^{j}_{m}$ are defined through the equation~$(\bar{f}^{j}_{m})^2=1-(f^{j}_{m})^2$. The addition of the two stochastic channels, $\bar{\mathscr F}+{\mathscr F}$, is trace-preserving, i.e., it describes a deterministic operation, with no post-selection. The final measurement is given by the seed~$\tilde\Omega$ defined after Eq.~(\ref{eq:stoch}) for both favorable and unfavorable events. Thus, we can easily compute ${\bar\sigma}^{2}(S)$  for the the latter, as well as~$\sigma_{{\rm all}}^{2}(S)$, where all outcomes are included. Clearly, we must have that $\sigma_{{\rm all}}^{2}(S)\geq \mse_{\rm det}$~\cite{combes_quantum_2014}, as~$\mse_{\rm det}$ refers to the optimal deterministic protocol.
\begin{figure}[htbp] %  figure placement: here, top, bottom, or page
   \centering
   \includegraphics[width=3.4in]{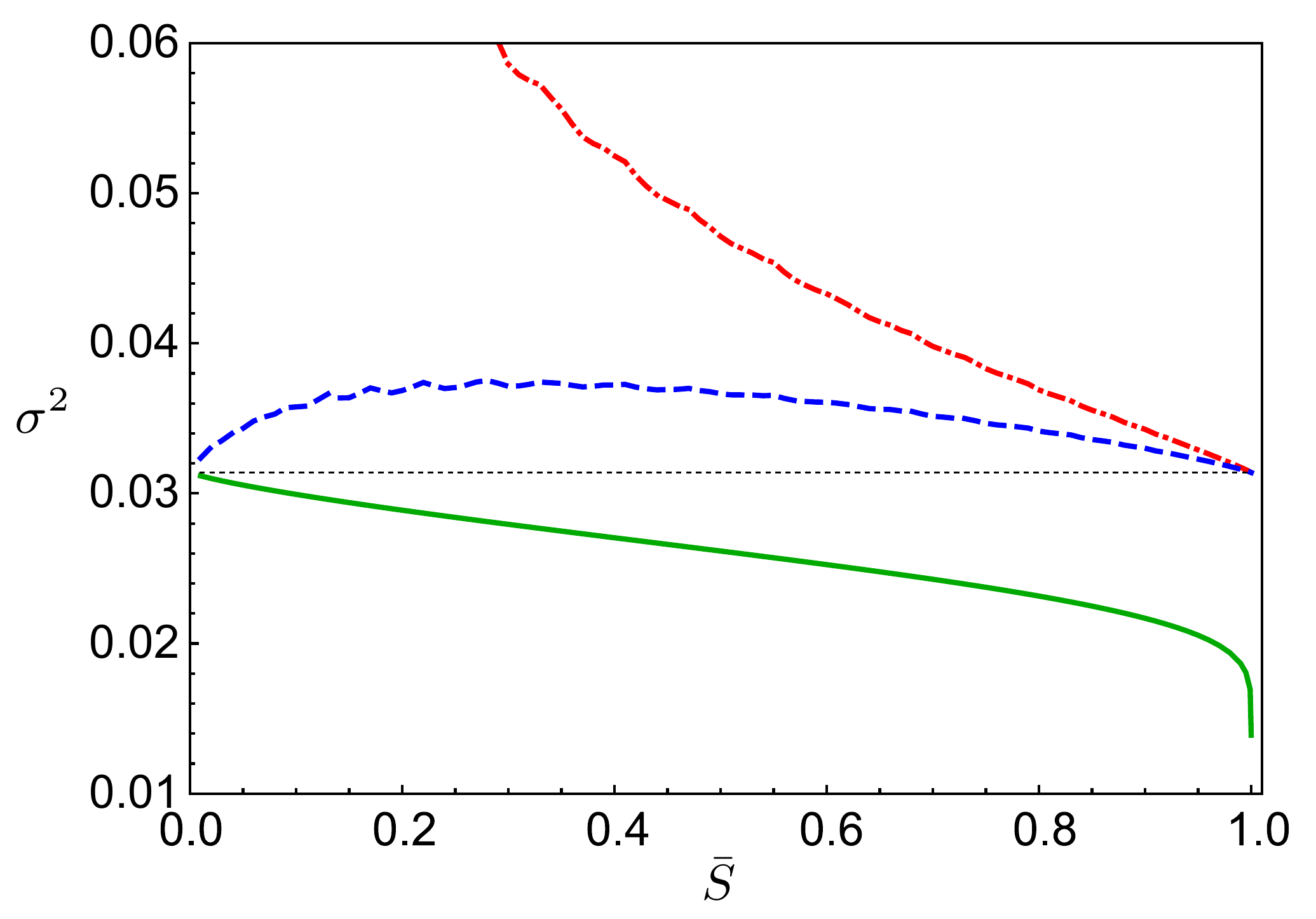} 
   \caption{{\bf Scavenging information.} Uncertainty $\sigma^2$ vs probability of abstention~$\bar S$ from numerical optimization for $n = 50$ and $r = 0.8$.  The green solid (red dash-dotted) correspond to $\sigma^2_{\rm opt}$ ($\bar\sigma^2$), where only the favorable (unfavorable) events are taken into account. The dashed curve corresponds to~$\sigma_{\rm all}^2$, where an estimate is provided on \emph{all} outcomes, favorable or unfavorable. For low success probability ($\bar S$ close to unity), both~$\bar\sigma^2$, and~$\sigma_{\rm all}^2$, approach the uncertainty of the deterministic protocol~$\sigma^2_{\rm det}$ (dotted line).}
   \label{fig:scav}
\end{figure}

As shown in Figure~\ref{fig:scav}, a protocol that is optimized for some probability of abstention $\bar{S}$, performs only slightly worse when forced to provide always a conclusive outcome. In particular, we notice that if such protocol is designed to work at the ultimate limit regime, with uncertainty~$\mse_{\mathrm{ult}}$, which requires a very large abstention probability ($S\to 0$)~\cite{combes_quantum_2014}, its  performance coincides with that of the optimal deterministic protocol.  Actually, this observation follows (see~\ref{app gent}) from Winter's gentle measurement lemma (Lemma~9 in~\cite{winter_coding_1999}), which states that a measurement with a highly unlikely outcome causes only a little disturbance to the measured quantum state. This is in contrast to the claims in~\cite{combes_quantum_2014}, where a random estimate is assigned to the discarded events.

\section{Discussion}

We have shown that abstention or post-selection can  counterbalance the adverse errors in a noisy metrology task. Our results are theoretical and concern abstract systems of $n$  qubits. However, they apply  to different quantum metrology implementations, ranging from  Ramsey interferometry for frequency standards \cite{bollinger_optimal_1996,huelga_improvement_1997}, atomic magnetometry~\mbox{\cite{budker_optical_2007,napolitano_interaction-based_2011}}, and quantum photonics (single or multi-mode setups), where the number operator introduced here will play the role of number of photons.

Post-selection is already widely used for preparing quantum information resources, e.g., single photons from weak coherent pulses, heralded down-conversion for EPR-type states, or NOON states for metrology applications. Although some degree of post-selection is common in experiments, its tailored optimised use is not fully exploited. Only recently there have been important developments  
in this direction in the context of weak value amplification \cite{dressel_colloquium:_2014,ferrie_weak_2014,hosten_observation_2008}. We note on passing that these schemes can be considered a particular instance of our general set-up, and hence are subject to our bounds.

The optimal probabilistic measurement presented here can be understood as a filtering process selecting the total angular momentum followed by a modulating filter, and a final standard covariant phase measurement. 
The latter can be implemented by the (almost) optimal adaptive scheme proposed in \cite{berry_optimal_2000,paris_2015}. The modulation could be implemented by sequential use of amplitude-damping channels taking inspiration from recent experiments in state amplification \cite{ferreyrol_implementation_2010,zavattaa._high-fidelity_2011,kocsis_heralded_2013}. In implementations that allow for an individual control of the qubits, such as ion traps, the projection onto the angular momentum basis can be efficiently carried out \cite{bacon_efficient_2006}. 
For implementations with less degree of control, the projection onto the fully symmetric sub-space can, as a last resort, be implemented by post-selecting outcomes with this symmetry. For instance a simple Stern Gerlach measurement could lead to outcomes ($m=J$) with a precision beyond the deterministic limits.

Regarding the implementation of our conjectured optimal probe state one can use available non-linear $N^{2}$-type two-body interactions to turn the multi-copy gaussian profile to the wider optimal gaussian. Although, our case-study focuses on local dephasing noise, 
our methods can be adapted and similar, if not greater, benefits are expected for more general and implementation-specific noise models, including correlated noise.

In conclusion, we have shown what are the ultimate limits in precision reachable by any (deterministic or stochastic) quantum metrology  protocol in a realistic scenario with local decoherence. We have derived the optimal bounds that can be reached when a certain rate of abstention is allowed and hence provided a full assessment of the risks and benefits of the probabilistic strategy.
The benefits are clear for finite and for  asymptotically large number of copies, and the  precision is strictly better than that attained by deterministic strategies, which include optimal preparation of probe states. The ultimate quantum metrology scaling limit is only reached with a large abstention rate. However, in that case we have shown that it is possible to obtain estimates with  standard (deterministic) precision from the discarded events. In this sense, seeking ultra-sensitive measurements is a low-risk endeavour.

\section{Acknowledgements}

This research was supported by the Spanish MINECO contract FIS2013-40627-P and the Generalitat de Catalunya CIRIT, contract 2014-SGR966.

\appendix 

%\section{Appendices}

\section{Notation \label{subs:notation}}
Throughout this section we use the following notation. The $n$-qubit computational basis is denoted by $\{| b\rangle\}_{b=0}^{2^n-1}$, where~$b=b_1b_2\cdots b_n$ is a binary sequence, i.e., $b_i=0,1$ for $i=1,2,\dots,n$. We denote  by $|b|$ the sum of the $n$ digits of $b$, i.e.,  $|b|:=\sum_{j=1}^n b_j$. The digit-wise sum of $b$ and $b'$ modulo~$2$ will be simply denoted by $b+b'$, hence~$|b+b'|$ can be understood  as the Hamming distance between $b$ and $b'$, both viewed as binary vectors.

The permutations of $n$ objects, i.e., the elements of the symmetric group $S_n$, are denoted by $\pi$. We define the action of a permutation $\pi$ over a binary list~$b$ as \mbox{$\pi(b):=b_{\pi(1)}b_{\pi(2)}\cdots b_{\pi(n)}$}. This induces a unitary representation of the symmetric group on the Hilbert space~${\cal H}^{\otimes n}$ of the $n$ qubits through the definition \mbox{$U_\pi|b\rangle:=|\pi(b)\rangle$}. The (fully) symmetric subspace of~${\cal H}^{\otimes n}$, which we denote by ${\cal H}^{\otimes n}_+$, plays an important role below. An orthornormal basis can be labelled $\beta=|b|$, where $\beta=0,1,\dots,n$:
\begin{equation}
|\beta\rangle%={1\over n!}\begin{pmatrix}n\\\sigma\end{pmatrix}^{1/2}\sum_{\pi\in{\cal S}_n}U_\pi|s\rangle
={n\choose \beta}^{-1/2}\sum_{b\in B_\beta}|b\rangle  \mbox{ with } \beta=0,1,\ldots n
\label{basis H_+}
\end{equation}
where $B_\beta=\{b : |b|=\beta\}$. It is well-known that the symmetric subspace~${\cal H}^{\otimes n}_+$ carries the irreducible representation of spin $j=J:=n/2$ of $SU(2)$. In this language, the magnetic number $m$ is related to~$\beta$ by $m=n/2-\beta$ (here we are mapping $\beta_i\to m_i=(-1)^{\beta_i}/2$ for qubit $i$). In other words, we map $|\beta\rangle\to|n/2,n/2-\beta\rangle$, where we stick to the standard notation $|j,m\rangle$ for the spin angular momentum eigenstates. 

We will be concerned with evolution under unitary transformations $U_\theta:=u^{\otimes n}_\theta$, where $u_\theta=\exp(i\theta |1\rangle\langle1|)$, $\theta\in(-\pi,\pi]$. The operator $N$ such that $U_\theta={\rm e}^{i\theta N}$  will be referred to as number operator for obvious reasons: $N|b\rangle=|b||b\rangle$. The effect of noise is taken care of by a CP map ${\mathscr D}$, so the actual evolution of an initial $n$-qubit state~$\psi:=|\psi\rangle\langle\psi|$ is $\psi\to {\mathscr D}(U_\theta\psi U^\dagger_\theta)=\rho_\theta$. 

With this notation the uncertainty and success probability in Eqs.~(\ref{F_max}) and~(\ref{succ}) can be written as
\begin{eqnarray}
\sigma^{2}({S})
&=&2-\frac{2}{S} \max_{\Omega} \sum_{b,b'} \Omega_{b,b'} \rho_{b',b}\delta_{|b'|,|b|+1},
\label{F(S) b}
\\
S&=&
 \sum_{b,b'} \Omega_{b,b'} \rho_{b',b}\delta_{|b'|,|b|},
 \label{S b}
\end{eqnarray}
where the Kronecker delta tensors result from the integration of~$\hat\theta$. 
\medskip

\section{ Local dephasing: Hadamard channel \label{subs:localdephasing}}
In this paper we consider uncorrelated dephasing noise, which can be modeled by phase-flip errors that occur with probability $p_f$. i.e., at the single qubit level, the effect of the noise is~$\varrho\to (1-p_f)\varrho+p_f\,\sigma_z\varrho\,\sigma_z$, where $\sigma_z$ is the standard Pauli matrix $\sigma_z={\rm diag}(1,-1)$. 
For states of~$n$ qubits, this, so called dephasing channel ${\mathscr D}$, is most easily characterized through its action on the operator basis~$\{|b\rangle\langle b'|\}_{b,b'=1}^{2^n-1}$ as 
\begin{equation}
{\mathscr D}(|b\rangle\langle b'|)=r^{|b+b'|}|b\rangle\langle b'|,
\end{equation}
where the parameter $r$  is related to the error probability~$p_f$ through $r=1-2p_f$. The effect of ${\mathscr D}$ on a general $n$-qubit state $\varrho=\sum_{b,b'}\varrho_{b,b'}|b\rangle\langle b'|$ can then be written as the Hadamard (or entrywise) product 
\begin{equation}
{\mathscr D}(\varrho)=\sum_{b,b'}r^{|b+b'|}\varrho_{b,b'}|b\rangle\langle b'|:={\cal D}\circ\varrho,
\label{Hadamard}
\end{equation}
where ${\cal D}:=\sum_{b,b'}r^{|b+b'|}|b\rangle\langle b'|$ and hereafter we understand that the sums over sequences run over all possible values of $b$ (and $b'$) unless otherwise specified. Note that Hadamard product is basis-dependent.

\medskip

\section{Symmetric probes. \label{subs:symmetricprobes}} If the probe state is fully symmetric, i.e., $|\psi\rangle\in{\cal H}^{\otimes n}_+$, it can be written as $|\psi\rangle=\sum_{\beta} \psi_\beta |\beta\rangle$, where~$|\beta\rangle$  is defined in Eq.~(\ref{basis H_+}) and the components are related to those in Eq.~(\ref{eq:symm}) by $c_m=\psi_{J-m}$ and can be taken to be positive with no loss of generality (any phase can be absorbed in the measurement operators). Then,~$\rho={\mathscr D}(\psi)={\cal D}\circ\psi$ in Eqs.~(\ref{F(S) b}) and~(\ref{S b}) becomes
\begin{equation}
\rho=\sum_{\beta,\beta'}{\psi_{\beta'} \psi_{\beta}\over {n\choose \beta}^{1/2}{n\choose \beta'}^{1/2} }\sum_{b\in B_\beta} \sum_{b'\in B_{\beta'}}\! r^{|b+b'|}|b'\rangle\langle b|  .
\end{equation}
Since $\rho$ is permutation invariant, $\Omega$ can be chosen to be so and we can easily write Eqs.~(\ref{F(S) b}) and~(\ref{S b}) in the  spin basis. We just need the non-zero Clebsch-Gordan matrix elements~$\langle j,m'|b'\rangle\langle b|j,m\rangle$, where implicitly $m=J-\beta$, $m'=J-\beta'$. If we introduce the shorthand notation ${\cal D}^j_{m',m}:=\langle j,m'|{\cal D}|j,m\rangle$, then using~\cite{calsamiglia_local_2010}, we have
\begin{eqnarray}
{\cal D}^j_{m',m}&=&
\sum_{b\in B_\beta} \sum_{b'\in B_{\beta'}}\! r^{|b+b'|}\langle j,m'|b'\rangle\langle b|j,m\rangle\nonumber \\
&=&(1-r^2)^{J-j}r^{m-m'} \sum_{k}[\Delta_k^{(j)}]_m^{m'} r^{2k} ,
\label{eq:CG}
\end{eqnarray}
where
\begin{equation}
\big[\Delta^{(j)}_k\big]^{m'}_m\!\!:=\!{\sqrt{(j\!-\!m)!(j\!+\!m)!(j\!-\!m')!(j\!+\!m')!}\over
(j\!-\!m\!-\!k)!(j\!+\!m'\!-\!k)!(m\!-\!m'\!+\!k)!k!
} ,
\label{coeff Wigner}
\end{equation}
and the sums run over all integer values for which the factorials make sense. Recalling Eq.~\eqref{eq:rhoj}, a simple expression, involving just a sum over $k$ in Eq. \eqref{coeff Wigner}, for \mbox{$\rho^j_{m,m'}=p_j^{-1} \mathrm{tr}(\ketbra{j,m}{j,m'}\otimes \id_j \; \rho)$} follows by combining the above results. In short,
\begin{equation}\label{eq:rhojmm}
\rho^j_{m',m}={1\over p_j}{c_{m'}c_m\over{n\choose J-m'}^{1/2}{n\choose J-m}^{1/2}} {\cal D}^j_{m',m},
\end{equation}
where
\begin{equation}\label{eq:pj}
p_j=\nu_{j}\sum_{m}{c_{m}^{2}\over{n\choose J-m}} {\cal D}^j_{m,m},
\end{equation}
and the multiplicity is given by,
\begin{equation}
\nu_j={n\choose J-j}{2j+1\over J+j+1}
\end{equation}
and $a^j_m$ in Eq.~(\ref{eq:H^j_mm}) becomes
\begin{equation}
a^j_m={{\cal D}^j_{m,m+1}\over\sqrt{{\cal D}^j_{m,m}{\cal D}^j_{m+1,m+1}}}\,.
\label{a^j_m D}
\end{equation}

\section{{Relevant expressions for the multi-copy state.}\label{subs:relevantExpres}} If the input state is of the form given in Eq.~\eqref{eq:multicopy}, the above expressions, Eqs.\eqref{eq:rhojmm} and \eqref{eq:pj}, become

\begin{equation}
\label{eq:rhojmm2}
\rho^{j}_{m',m}={{\cal D}_{m',m}^j\over \sum_{m}{\cal D}_{m,m}^j}  \mbox{ and } p_j=\nu_j\;2^{-n}\sum_{m}{\cal D}_{m,m}^j\,,
\end{equation}
where
\begin{equation}
\label{eq:sumD}
\sum_{m}{\cal D}_{m,m}^j=(1-r^2)^{J-j}{(1+r)^{2j+1}-(1-r)^{2j+1}\over 2\;r}
\end{equation}

The probability to find the state in the fully symmetric subspace ($j=J$) is important when assessing the success probability of the the ultimate bounds.  Since the multiplicity for the maximum  spin $J$ is equal to one, it can be readily seen that $p_{J}$ scales as
\begin{equation}
p_J\sim {\rm e}^{-n[\log 2-\log(1+r)]}\,,
\end{equation}

The critical probability $s_{j}^{*}$ within a block can also be computed in the asymptotic limit $j\gg 1$ from  Eq.~\eqref{eq:Scrit} 
\begin{equation}
s_j^*={\rho_{j,j}^j\over (\xi_j^j)^2}\sim  \expo{-2j\log ({1+r})}
\end{equation}
where $\xi_m^j$ is the gaussian ground state, with $(\xi_j^j)^2\sim \exp(-(1-r^2)\sqrt{j/4r})$. For $m=m'=j$  Eq.~\eqref{eq:CG} gives $D^j_{j,j}=(1-r^2)^{J-j}$ which together with  Eqs.~\eqref{eq:rhojmm2} and Eq.\eqref{eq:sumD} gives  $\rho_{j,j}^j\sim \exp[ 2 j \log(r+1) ]$. This scaling dominates over that of $\xi_m^j$ , and hence determines the scaling of $s_{j}^{*}$.
From here we obtain critical value for the overall success probability $S^*=p_{J}\;s_{J}^*\sim\expo{-n\log\;2}$.

\medskip
\section{{Ultimate bound without in-block filtering.}\label{subs:ultimateBwithout}}
In the results section we discuss the possibility to prepare a probe state such that after the action of  noise becomes an optimal state within  the fully symmetric subspace $j=J$. Here we give what its critical success probability, which only entails computing $p_{J}$.

For this purpose we first recall that  the optimal filtered state $\xi^j_m$,  defined  before Eq.~(\ref{eq:H^j_mm}),  has to fulfil 
\begin{equation}
\xi^j_m=f^j_m c_m \sqrt{{\nu_j {\cal D}^j_{m,m}\over s_j p_j {n\choose J-m}}}\  .
\label{eq:xi expl}
\end{equation}
The probability of falling in the block of maximum  spin $J$ for a given filtered state~$\xi^j$ can be easily derived from Eq.~(\ref{eq:xi expl}) recalling that the probe state $\psi$ is normalized, and thus $\sum_{m}c_m^2=1$. Solving Eq.~(\ref{eq:xi expl}) for $c^2_m/p_J$ and summing over $m$ we obtain
\begin{equation}
{1\over p_J}=s_J\sum_{m=-J}^{J}{{n\choose J-m}\over{\cal D}^{J}_{m,m}}\left({\xi^J_m\over f^J_m}\right)^2.
\end{equation}
Now, for our strategy  all $j$ but the maximum one, \mbox{$j=J$}, are filtered out, and no further filtering is required within the block $J$, i.e. we have $f^J_m=1$, for all $2J+1$ values of $m$. Then $s_J=1$ and
\begin{equation}
p_J=\left\{\sum_{m=-J}^J {{n\choose J-m}\over{\cal D}^{J}_{m,m}}\left({\xi^J_m}\right)^2\right\}^{-1}.
\end{equation}
In the asymptotic limit the probability $p_J$ can be estimated by noticing that the optimal distribution $(\xi^j_m)^2$ is much wider than ${n\choose J-m}/{\cal D}^{J}_{m,m}$ and can be replaced by~$(\xi^J_0)^2$. Around $m=0$, we can use the asymptotic formulas
\begin{eqnarray}
{\cal D}^{j}_{m,m}&\sim& (1-r^2)^{J-j}{(1+r)^{2j+1}\over 2\sqrt{\pi r j}}{\rm e}^{-r m^2/j},\label{D^j_mm Gauss}\\[.5em]
{n\choose J-m}&\sim&{2^n\over\sqrt{\pi J}}{\rm e}^{-m^2/J}.
\end{eqnarray}
They can be derived using the Stirling approximation and saddle point techniques. Eq.~(\ref{D^j_mm Gauss}) also requires the Euler–Maclaurin approximation to turn the sum over $k$ in Eq.~(\ref{eq:CG})  into an integral that can be evaluated using again the saddle point approximation. Retaining only exponential terms, $S^{*}=p_J\sim(1+r)^n/2^n={\rm e}^{-n[\log 2-\log(1+r)]}$. 
\medskip

\section{Equivalence of worst-case and expected loss\label{sect:wc=bay} } 
Here we give a simple proof that for phase estimation, and assuming a covariant signal, such as $\rho_\theta={\mathscr D}(U_\theta \psi U_\theta^\dagger)$, and a flat prior $\pi(\theta)=1/(2\pi)$ (see Sec III.A),  the worst-case loss in Eq.~(\ref{Losswc}),
\begin{equation}
\label{fidwc}
L_{\mathrm{wc}}   =  \sup_{\theta\in (-\pi,\pi] }  \int  {\rm d} \hat \theta  \,    p(\hat \theta | \theta , \mathrm{succ})\,  \ml(\theta,\hat\theta),
\end{equation}
 and the expected loss in Eq.~\eqref{fidelity-def},
\begin{equation}
L_{\rm s}=\frac{1}{S}\int\!\! \int\! {\rm d} \hat\theta\,{\rm d}\theta\, p(\theta,\hat\theta, \mathrm{succ}) \ml(\theta,\hat\theta),
 \label{Fav}
\end{equation}
take the same value, and so do the corresponding success probabilities. 
%We recall that $\ml_{\rm p}(\theta,\hat\theta)=4\sin^2[(\theta-\hat\theta)/2]$ and note that we have assumed a flat prior probability  $\pi(\theta)=1/(2\pi)$ for $\rho_\theta=U_\theta\rho U_\theta^\dagger$. %The proof can be easily extended to any covariant family of states. 

We first rewrite Eq.~\eqref{Fav} as
\begin{eqnarray}
L_{\rm s}&=&\frac{1}{S}\int\!\! \int\! {\rm d} \hat\theta\,{\rm d}\theta\, p(\hat\theta | \mathrm{succ}, \theta)p(\mathrm{succ}, \theta) \ml(\theta,\hat\theta)= \nonumber\\
&=&\int\!{\rm d}\theta\,p(\theta|\mathrm{succ})\!\int\! {\rm d}\hat\theta\, p(\hat\theta|\theta, \mathrm{succ}) \ml(\theta,\hat\theta) ,
\label{Favp}
\end{eqnarray}
where in the second equality we have used the fact that $p(\mathrm{succ}, \theta)=p(\theta|\mathrm{succ}) S$.
When written in this form, we note the analogy between $L_{\rm s}$ in Eq.~\eqref{Favp} and $L_{\mathrm{wc}}$ in Eq.~\eqref{fidwc}, where the average over $p(\theta|\mathrm{succ})$ is replaced by the supremum over~$\theta$. Because of this, we obviously have~$L_{\rm s}\le L_{\rm wc}$. We just need to show that the opposite inequality also~holds.

We remind the reader that the minimum expected loss,~$L_{\rm s}$, can be attained by a covariant measurement, ${\mathscr M}=\{M_{\hat\theta}=U_{\hat\theta}\Omega U^\dagger_{\hat\theta}/(2\pi)\}_{\hat\theta\in(-\pi,\pi)]}$, where~$\Omega$  is a suitable seed  (the optimal seed). 
Because of covariance, we note that for any phase $\theta'$ we have $p(\hat\theta|\theta,{\rm succ})=p(\hat\theta+\Delta\theta|\theta',{\rm succ})$, where  \mbox{$\Delta\theta=\theta'-\theta$}. Likewise, we have $\ml_{\rm p}(\theta,\hat\theta)=\ml_{\rm p}(\theta',\hat\theta+\Delta\theta)$.
By shifting variables $\hat\theta+\Delta\theta\to \hat\theta$ in Eq.~(\ref{Fav}), the integrant becomes independent of the variable $\theta$, which can be trivially integrated  to give 
\begin{equation}
L_{\rm s}= \int d\hat\theta \,p(\hat\theta|\theta',{\rm succ}) \ml_{\rm p}(\theta',\hat\theta)
\end{equation}
for any $\theta'$. It follows that the worst case loss for this particular measurement, $L^{\mathscr M}_{\rm wc}$, satisfies $L^{\mathscr M}_{\rm wc}=L_{\rm s}$. Since~${\mathscr M}$ needs not to be the measurement that minimizes Eq.~(\ref{fidwc}), we have $L_{\rm wc}\le L^{\mathscr M}_{\rm wc}=L_{\rm s}$.
%%
%\begin{equation}
%L_{\rm wc}=\sup_{\theta\in(-\pi,\pi]} \int d\hat\theta \,p(\hat\theta|\theta,{\rm succ}) \ml_{s}(\theta,\hat\theta)\leq L_{\rm s},
%\end{equation}
%%
%where the last inequality states that the measurement that mimizes  $L_{\rm wc}$ need not to minimize the expected loss $L_{\mathrm wc}$. 
Combining this with the opposite inequality, derived after Eq.~(\ref{Favp}), we conclude that $L_{\rm s}=L_{\mathrm wc}$.

Proceeding along the same lines, we note that
\begin{equation}
S=\int {d\theta\over2\pi} \int d\hat\theta \,p(\hat\theta|\theta,{\rm succ}) 
=\int d\hat\theta \,p(\hat\theta|\theta',{\rm succ})
\end{equation}
for any $\theta'$. Hence $S=S_{\mathrm{wc}}$.

\medskip

\section{ The continuum limit: Particle in a potential box.\label{subs:continuum}}
Proceeding as in~\cite{gendra_quantum_2013,gendra_optimal_2013, summy_phase_1990}, one can easily derive from Eqs.~(\ref{eq:fidj}), (\ref{eq:H^j_mm}) and~(\ref{eq:fidjmax}) the following equation:
\begin{eqnarray}
\langle\xi^j|H^j\!|\xi^j\rangle&=&\!\sum_{m=-j}^{j-1} \!\!\left\{\!\left(\xi_{m+1}\!-\!\xi_m\right)^2\!\!+{V^j_m\over j^2}\xi^2_m\!\right\}\nonumber\\
&+& a_{-j}\xi^2_{-j}\!+a_j\xi^2_{j},
\label{eq:discrete}
\end{eqnarray}

where $V^j_m=2j^2(1-a_m)$ and we have dropped the superscript $j$ in $\xi^j_m$ to simplify the expression. In the asymptotic limit, as $j$ becomes very large, $m/j=x$ approaches a continuum variable that takes values in the interval~$[-1,1]$. Accordingly, the values $\{\sqrt j \xi_m\}$ approach a real function that we denote by $\varphi(x)$. With this, the former equation becomes 
\begin{eqnarray}
\label{eq: ?}
 \langle\xi^j|H^j\!|\xi^j\rangle\!&=&{
1\over j^2}%
%{\varphi_j(-1)^{2}+\varphi_{j}(1)^{2}\over j}\right.\nonumber\\[.5em]
%&+&\left.
\int_{-1}^1 \!\!dx \left\{  \left[{d\varphi(x)\over dx}\right]^2\!\!+{V^j(x)}\varphi(x)^{2}\right\},\nonumber\\[.5em]
:\!&=&{
1\over j^2}\bra{\varphi} {\cal H}^j\ket{\varphi},
\end{eqnarray}
where
\begin{equation}
V^j(x)= 2 j^2 (1-a^j_{xj}),\quad {\cal H}^j:=-d^2/dx^2+V^j(x),
\end{equation}
and we have dropped the boundary term $[\varphi^2(-1)+\varphi^2(-1)]/j$ that stems from the second line in Eq.~(\ref{eq:discrete}). Minimization of~$\langle\xi^j|H^j\!|\xi^j\rangle$ require the vanishing of this term, and Eq.~(\ref{eq: sigmaj2}) readily follows.
The formula
\begin{equation}
V^j(x)= j{1-r^2\over2r\sqrt{1-(1-r^2)x^2}} 
\end{equation}
[also in Eq.~(\ref{eq:pot})] follows from the asymptotic expression of~$a^j_m$, defined in Eq.~(\ref{eq:H^j_mm}). Our starting point are Eqs.~(\ref{a^j_m D}) and~(\ref{eq:CG}). The sum over $k$ in the latter can be evaluated using the Euler–Maclaurin formula and the saddle point approximation.

\medskip

\section{ Symmetric probe is optimal and no benefit in probe-ancilla entanglement.}
We next show that permutation invariance enables us to choose with  no loss of generality the probe state  $|\psi\rangle$ from the symmetric subspace ${\cal H}^{\otimes n}_+$ and the seed~$\Omega$ to be fully symmetric. 

We first write Eqs.~(\ref{F(S) b}) and~(\ref{S b}) in a more compact form. We define $\Delta$  through the relation $\sigma^2(S)=(1+S^{-1}\min_{\psi,\Omega}\Delta)/2$, where the maximization is performed also over the probe states since here we are concerned with the ultimate precision bound. We also introduce a slight modification of ${\cal D}$ that includes the Kronecker delta tensor:  ${\cal D}_l:=\sum_{b,b'} r^{|b+b'|}\delta_{|b'|,|b|+l}|b\rangle\langle b'|$, $l=0,1$. Then,
\begin{equation}
\Delta={\rm tr}\left[(\psi\circ \Omega) {\cal D}_1\right],\quad
S={\rm tr}\left[(\psi\circ \Omega) {\cal D}_0\right],
\label{Delta S}
\end{equation}
where we have used that ${\rm tr}[A(B\circ C)]={\rm tr}[(C\circ A)B]$ if $B=B^t$.
The result we wish to show follows from the invariance of the noise under permutations of the $n$ qubits, namely, from $U_\pi {\cal D}_lU^\dagger_\pi={\cal D}_l$, for any $\pi\in {S}_n$, which implies that the very same value of $\Delta$ and $S$ attained by some given measurement seed $\Omega$ and some  initial state $\psi$, i.e., attained by $\psi\circ\Omega$, can also be attained by $U_\pi( \psi\circ\Omega) U^\dagger_\pi$, and likewise by the average $(n!)^{-1}\sum_{\pi\in {S}_n}U_{\pi}( \psi\circ\Omega) U^\dagger_{\pi}$.

The proof starts with yet a few more definitions: given a fully general probe state  $|\psi\rangle$, we define the $n+1$ normalized states %$|\xi_\sigma\rangle=\sum_{s\in S_\sigma}\xi^\sigma_s|s\rangle$
$|\phi_\beta\rangle=\sum_{b\in B_\beta}(\psi_b/\psi_\beta)|b\rangle$, $\beta=0,1,\dots, n$ where %$\xi^\sigma_s=\left(\sum_{s\in S_\sigma}|\psi_s|^2\right)^{-1/2} \psi_s $
\mbox{$\psi^2_\beta=\sum_{b\in B_\beta}|\psi_b|^2$}, and write
$
|\psi\rangle=\sum_{\beta=0}^n \psi_\beta |\phi_\beta\rangle
$. 
Additionally, we define
\begin{equation}
|\phi\rangle=\sum_{\beta=0}^n
{n\choose\beta}^{1/2}
|\phi_\beta\rangle, \quad \phi=|\phi\rangle\langle\phi|  .
\end{equation}
We obviously have $\phi\circ\Omega\ge0$, as the Hadamard product of two positive operators is also a positive operator, and
\mbox{$
(n!)^{-1}\sum_{\pi\in{S}_n} U_\pi \left(\phi\circ\Omega\right) U^\dagger_\pi\ge 0
$}, as this expression is a convex combination of positive operators.
Similarly, the seed condition~$\openone-\Omega\ge0$ implies $(n!)^{-1}\sum_{\pi\in{S}_n} U_\pi \left[\phi\circ\left(\openone-\Omega\right)\right] U^\dagger_\pi\ge 0$. But
\begin{equation}
{1\over n!}\!\!\sum_{\pi\in{S}_n} \!\!U_\pi\! \left(\phi\circ \openone\right)\! U^\dagger_\pi
\!=\!\sum_{\beta=0}^n {{n\choose\beta}\over n!}\!
\left(
\sum_{\pi\in{S}_n} \!U_\pi\phi_\beta U^\dagger_\pi  \right)   \circ \openone,
\label{<1a}
\end{equation}
since the diagonal entries of $\phi$ and $\phi_\beta$ transform among themselves under permutations. 
The right hand side can be written as
\begin{equation}
\kern-.25em
\sum_{\beta=0}^n\sum_{b\in B_\beta}\!\!
%\left(
 {{n \choose \beta}\over n!}\!\!\sum_{\pi\in{S}_n}\!\! \!{|\psi_{\pi^{-1}\!(b)}|^2\kern-.3em\over\psi^2_\beta} |b\rangle\langle b|%\circ
%\right) 
%\openone\right)
\!=\!
\sum_{\beta=0}^n\sum_{b\in B_\beta} \!\!%\left(\sum_{s'\in S_\sigma} {|\psi_{s'}|^2\over\psi^2_\sigma}
% \right) 
%\right)
|b\rangle\langle b|
=\openone,
\label{<1}
\end{equation}
where we have used that,  for any  $b\in B_\beta$, the set $\{\pi^{-1}(b)\}_{\pi\in{S}_n}$ contains exactly~$\beta!(n-\beta)!$ times each one of the elements of $B_\beta$. 
It follows from Eqs.~(\ref{<1a})  and~(\ref{<1}) that $ \Omega^{\rm sym}:=(n!)^{-1}\sum_{\pi\in{S}_n} U_\pi \left(\phi\circ\Omega\right) U^\dagger_\pi$ satisfies
$0\le \Omega^{\rm sym}\le \openone$ and is invariant under permutations of the $n$ qubits. It is, therefore, a legitimate fully symmetric measurement seed. Moreover,
\begin{equation}
{1\over n!}\!\!\sum_{\pi\in{S}_n}\!\!U_\pi\!\left(\psi\!\circ\!\Omega\right)\!U^\dagger_\pi
\!=\!
\sum_{\beta,\beta'} {\psi_\beta\psi_{\beta'}\over  {n \choose \beta}^{1/2} {n \choose \beta'}^{1/2}}
\openone_\beta
\Omega^{\rm sym}
 \openone_{\beta'},
 \label{symm str}
 \end{equation}
where $\openone_\beta$ is the projector into the subspace with $|b|=\beta$, namely $\openone_\beta:=\sum_{b\in B_\beta}|b\rangle\langle b|$. Thus, recalling the definition of $|\beta\rangle$ in Eq.~(\ref{basis H_+}), the righthand side of Eq.~(\ref{symm str}) can be readily written as
\begin{equation}
\Bigg(\sum_{\beta,\beta'} \psi_\beta\psi_{\beta'}|\beta\rangle\langle\beta'|\Bigg)\circ \Omega^{\rm sym}
=\psi^{\rm sym}\circ \Omega^{\rm sym} ,
\end{equation}
where~\mbox{$|\psi^{\rm sym}\rangle:=\sum_{\beta=0}^n\psi_\beta |\beta\rangle\in{\cal H}^{\otimes n}_+$}. It follows from these results and Eq.~(\ref{Delta S}) that the very same uncertainty and success probability attained by any pair $(|\psi\rangle,\Omega)$ of probe state and measurement seed is also attained by the state $|\psi^{\rm sym}\rangle\in{\cal H}^{\otimes n}_+$ and the fully symmetric seed $\Omega^{\rm sym}$. This completes the proof.

Now that we have learned that no boost in performance can be achieved by considering probe states more general than those in the symmetric subspace~${\cal H}^{\otimes n}_+$ (in the subspace of maximum  spin $j=J$),
we may wonder if entangling the probe with some ancillary system could enhance the precision. 
Here we show that this possibility can be immediately ruled out, thus extending the generality of our result. For this purpose we take the general probe-ancilla state $\ket{\Psi_{\rm PA}}=\sum_{b} \psi_{b} \ket{b}\ket{\chi_{b}}$,
where $\ket{\chi_{b}}$ are normalized states (not necessarily orthogonal) of the ancillary system.
The action of the phase evolution and noise on the probe leads to a state of the form \mbox{$\rho_{\rm PA}(\theta)\;=\;\sum_{b,b'} r^{|b+b'|} \;\expo{i \theta(|b|-|b'|)} \;\psi_{b}\psi_{b'} \ketbra{b}{b'}\otimes\ketbra{\chi_{b}}{\chi_{b'}}%=\sum_{b,b'} r^{|b+b'|} \expo{i \theta(|b|-|b'|)} c_{b}c_{b'} \ketbra{b}{b'}\otimes\ketbra{\phi_{b}}{\phi_{b'}}
$}.
This state could as well be prepared without the need of an ancillary system by taking instead an initial  probe state $\ket{\psi}=\sum_{b}\psi_{b}\ket{b}$ and performing the trace-preserving completely positive map defined by~$\ket{b}\to\ket{b}\ket{\chi_{b}}$ before implementing the measurement. This map can, of course, be interpreted as part of the measurement. It would correspond to a particular Neumark dilation of some measurement performed on the probe system alone, and hence it is included in our analysis.

\medskip

\section{ Scavenging at the ultimate precision limit.}\label{app gent}
The gentle measurement lemma~\cite{winter_coding_1999} states that if a measurement outcome occurs with very high probability, then the corresponding conditioned state is hardly disturbed. For concreteness, let us assume that some unfavorable event in a probabilistic protocol happens with probability~$\bar{S}=\mathrm{tr}[\bar {\mathscr F}(\rho_{\theta})]=1-\epsilon$, then, conditioned to this event, we have $\parallel\rho_{\theta}-\bar\rho_{\theta}\parallel_{1}\leq \sqrt{2}\epsilon$, where~$\bar{\rho}={\bar{\mathscr F}}(\rho_{\theta})/{\bar S}$.

Indeed, from  Eq.~\eqref{F_max} or \eqref{F(S) b} , we find that  the uncertainty of the scavenged events, $\bar \sigma^2(S)$, rapidly approaches that of the optimal deterministic machine $\sigma^{2}(S=0)$:
\begin{eqnarray}
&&\sigma^{2}(0)\!-\!\bar \sigma^{2}(S)\!=
\!\!\!\max_{0\leq \bar\Omega\leq \id}\!\! \tr{\bar W\!\bar\rho_{\theta}}-\!\!\!\max_{0\leq \Omega\leq \id}\!\! \tr{W\!\rho_{\theta}}\nonumber\\
&&\leq\max_{0\leq \Omega\leq \id} \mathrm{tr}[W(\rho_{\theta}\!-\!\bar\rho_{\theta})]\leq \parallel\!\rho_{\theta}\!-\!\bar\rho_{\theta}\!\parallel_{1}
\leq\! \sqrt{2}{S},
\label{eq:gentle}
%&&\sigma^{2}(0)\!-\!\bar \sigma^{2}(S)\!=\!\!\!\max_{0\leq \Omega\leq \id}\!\! \tr{W\!\rho_{\theta}}-\!\!\!\max_{0\leq \bar\Omega\leq \id}\!\! \tr{\bar W\!\bar\rho_{\theta}}\nonumber\\
%&&\leq\min_{0\leq \Omega\leq \id} \mathrm{tr}[W(\rho_{\theta}\!-\!\bar\rho_{\theta})]\leq \parallel\!\rho_{\theta}\!-\!\bar\rho_{\theta}\!\parallel_{1}
%\leq\! \sqrt{2}{S},
%\label{eq:gentle}
\end{eqnarray}
where $W$ (likewise $\bar W$) is shorthand for the matrix with entries $W_{b,b'}=2\Omega_{b,b'}\delta_{|b|,|b'|+1}$. 

We recall that, as $\sigma^2(S)$ approaches the ultimate bound $\sigma^{2}_{\rm ult}=1-(1-r^2)/(4nr)$, the success probability $S$ decreases exponentially. Eq.~(\ref{eq:gentle}) thus shows that such likely failure is not ruinous since 
in that event one can still attain the deterministic bound, i.e.,  one has \mbox{$\bar{\sigma}^{2}=\sigma^{2}(0)=1-(1-r^2)/(4nr^2)$}.

\end{document}